\documentclass[twocolumn,floatfix,showpacs,prd,aps,eqsecnum,tightenlines]{revtex4}
\usepackage{graphicx}
\usepackage{dcolumn}% Align table columns on decimal point
\usepackage{bm}% bold math
\newcommand{\includemygraphics}[2][1]{\includegraphics[#1]{figs/#2}} 

\begin{document}

%\preprint{UTBRG-2004-002, gr-qc/04mmnnn}

\title{Numerical integration of the Teukolsky Equation in the time domain}

\author{Enrique Pazos-\'Avalos\footnote{On leave from the Universidad
de San Carlos de Guatemala.} and Carlos O. Lousto}
\affiliation{Department of Physics and Astronomy,
and Center for Gravitational Wave Astronomy,
The University of Texas at Brownsville, Brownsville, Texas 78520}

\date{\today}

\begin{abstract}
We present a fourth order convergent $(2+1)$ numerical code to solve
the Teukolsky equation in the time domain. Our approach is to rewrite
the Teukolsky equation as a system of first order differential
equations. In this way we get a system that has the form of an
advection equation. This is used in combination with a series
expansion of the solution in powers of time. To obtain a fourth order
scheme we kept terms up to fourth derivative in time and use the
advection-like system of differential equations to substitute the
temporal derivatives by spatial derivatives. A local stability study
leads to a Courant factor of $1.5$ for the nonrotating case.  This
scheme is used to evolve gravitational perturbations in Schwarzschild
and Kerr backgrounds. Our numerical method proved to be fourth order
convergent in $r^*$ and $\theta$ directions. The correct power-law
tail, $\sim1/t^{2\ell+3}$, for general initial data, and
$\sim1/t^{2\ell+4}$, for time symmetric data, was found in the
simulations where the duration in time of the tail phase was long
enough.  We verified that it is crucial to resolve accurately the
angular dependence of the mode at late times in order to obtain these
values of the exponents in the power-law decay.  In other cases, when
the decay was too fast and round-off error was reached before a tail
was developed, the quasinormal modes frequencies provided a test to
determine the validity of our code.
\end{abstract}


\maketitle

\newcommand{\Dt}{\,\delta t}
\newcommand{\Dx}{\,\delta x}
\newcommand{\Dr}{\,\delta r}
\newcommand{\Drs}{\,\delta r^*}
\newcommand{\Dth}{\,\delta \theta}
\newcommand{\der}{\frac{d^5 u(\xi)}{dx^5}}
\newcommand{\bla}{\;\;}
\newcommand{\bbla}{\;\;\;\;}
\newcommand{\bfe}{{\bf e}}
\newcommand{\bfl}{{\bf l}}
\newcommand{\bfn}{{\bf n}}
\newcommand{\bfm}{{\bf m}}
\newcommand{\bfu}{{\bf u}}
\newcommand{\bfM}{{\bf M}}
\newcommand{\bfS}{{\bf S}}
\newcommand{\bfA}{{\bf A}}
\newcommand{\bfL}{{\bf L}}
\newcommand{\bfs}{{\bf s}}
\newcommand{\bfB}{{\bf B}}

%%%%%%%%%%%%%%%%%%%%%%%%%%%%%%%%%%%%%%%%%%%%%%%%%%%%%%%%%%%%%%%%%%%%%%%%
\section{Introduction}\label{Sec:Intro}

In recent years, there has been an increasing interest in numerically
solving General Relativity's (GR) field equations to provide an
accurate description of the gravitational radiation generated in
different astrophysical scenarios. This is due to fact that direct
measure of gravitational waves will soon be possible with large
interferometers such as LIGO and LISA. The astrophysical events that
produce gravitational waves are among the most energetic phenomena
ever seen. The best candidate for such observations are the collisions
of binary black hole systems. In order to detect this gravitational
radiation, accurate templates are needed, and this implies to solve
the non-linear GR equations. This proved to be a very challenging
task. Although in the last orbital stages (merger) of a binary black
hole collision it is necessary to solve Einstein's equations (full
numerical approach), the very last part of the coalescence (close
limit) the system can be considered as a single distorted black hole
and can be treated with perturbation theory. Comparisons between full
numerical simulations and perturbative methods show a surprising
agreement. This has encouraged people to go beyond matching close
limit and full numerical simulations in tandem to produce simulations
that neither of each technique alone was able to do. The general
method of coupling full numerical and approximate techniques is
the main development of the 'Lazarus 
project'~\cite{Baker:2000zh,Baker:2001nu,Baker:2001sf,Baker:2002qf,Baker:2004wv}.

Moreover, there are physical scenarios in which the metric departure
from the known static black hole solutions is always small outside the
horizon. Some examples are a particle (compact object) orbiting a
black hole, the propagation of gravitational waves and accretion disks
near black holes. The equations evolved in the perturbative method are
linear which is an advantage over the non-linear set of ten coupled
general relativistic equations.

One application of perturbation theory is the propagation of waves in
a curved spacetime. In this work, we will focus our study in the late
time behavior of a gravitational wave in such curved
spacetimes. Although the problem is not new, according to Andersson
\cite{andersson96} there are still some aspects for which there is no
definite answer or no answers at all; like the role played by highly
damped modes and the intermediate behavior of the power-law
tails. Power-law tails is the name given to the last phase in the
propagation of a wave in a curved background, as we will see
below. The general features of the evolution of such a test field in
the proximity of a black hole, as seen by a distant observer, can be
divided in three stages: i) Radiation emitted directly by the
perturbation source. It depends on the form of the initial field
(initial data). ii) Quasinormal ringing. It depends on the parameters
of the black hole. They are exponentially damped oscillations and
carry part of the gravitational radiated energy in astrophysical
processes like gravitational collapses. Quasinormal modes are
characterized by complex frequencies $\sigma$. Such modes can be
represented as $e^{i\sigma t}$, the real part of $\sigma$ is the
oscillation frequency and the imaginary part is the exponential
damping. iii) Power-law tail. The field decays with time according the
a power-law at very late times.

Most of the work on the late time radiative fall-off has been done in
spherically symmetric spacetime. In this case the solution admits a
decomposition in spherical harmonics. The field equations are reduced
to a wave equation with a effective potential. For the Schwarzschild
case this wave equation is
\begin{equation}
\left[ -\frac{\partial^2}{\partial t^2} + \frac{\partial^2}{\partial r^{*2}} - V(r) \right] \Psi_l(t,r) = 0
\label{eq:wavepot}
\end{equation}
where $r^*$ is the tortoise coordinate
\begin{equation}
r^* = r + 2M \ln(r/2M-1),
\label{eq:tortoise}
\end{equation}
and
\begin{equation}
V(r) = \left(1-\frac{2M}{r}\right)\left[ \frac{l(l+1)}{r^2} + \frac{2M}{r^3} \right]
\end{equation}
being $M$ the mass of the black hole. 

In 1972, Price \cite{price72} treated Eq.\ (\ref{eq:wavepot}) as a
perturbative expansion in powers of $M$. He showed that the
perturbation decays in time according to $t^{-(2l+3)}$ for $t \gg r
\gg M$. This situation occurs at a finite value of $r$ while
$t\to\infty$, which is called {\it timelike infinity}.

%Price \& Sun 1988, Gundlach, Price, Pullin 1994,and Ching,et. al. 1994 
Quasinormal modes and power-law tails have been studied by Leaver
\cite{leaver86}, who analyzed the problem in the frequency domain. He
found the correct late time behavior with a low-frequency
approximation. Quasinormal modes can be considered as the ``pure
tones'' of the black hole. Once excited, their damping and oscillation
frequency depend only on the parameters of the black hole. Ching
et. al. \cite{ching95} argue that the late time decay of the field can
be seen as a scattering due to the spacetime curvature. This implies
that the power-law behavior depends only on the asymptotic conditions
of the metric. In a recent work, Poisson \cite{poisson02} found that
in a weakly curved spacetime the late time dynamics is insensitive to
the non spherical aspects of the metric, being entirely determined by
the spacetime total mass.

In Ref.~\cite{Karkowski:2003et} it is found that the power-law tails are not an
universal phenomenon as thought, but the fall-off power depend on the
initial profile of the fields. While for the generic data with
$\Phi_\ell|^{t=0}\not=0$ and $\partial_t\Phi_\ell|^{t=0}\not=0$ the
predicted~\cite{price72} decay goes like $1/t^{2\ell+3}$, when you
start with an initially static field, i.e. $\Phi_\ell|^{t=0}\not=0$ and
$\partial_t\Phi_\ell|^{t=0}=0$ the predicted decay goes like $1/t^{2\ell+4}$.

Most of the work with the Teukolsky equation has been performed in the
frequency domain (quasinormal modes, wave scattering, motion of test
particles). Perturbation on the frequency domain can be reduced
analytically to solve ordinary differential equations and can lead to
a better understanding of the physics involved in the phenomena. Such
information is much more difficult to obtain from purely numerical
computations. However, in more complete treatments, the number of
frequencies that one needs to consider is orders of magnitude larger
than the number of points needed to resolve the $\theta$
direction. Furthermore, the study of quasinormal modes would require
higher resolution near $\omega=0$ to resolve the tails. The resolution
of the quasinormal modes is also sensitive to the spacing in
frequencies. These are the argument presented by Krivan
et. al. \cite{krivan97} in favor of a numerical treatment of
perturbations using the Teukolsky equation in the time domain. Another
motivation to work in the time domain is that when one tries to find a
solution of the radial Teukolsky equation in the frequency domain with
a source term that extends to infinity, the result is divergent. This
means that the Teukolsky equation needs to be regularized when sources
are present \cite{Poisson:1996ya,campanelli97}.

The difficulty with numerical integrations of the 
Teukolsky equation is the linear term in $s$ on
the first time derivative. Depending on the relative sign between this
term and the second time derivative, it can act as a damping or
anti-damping term. As described in \cite{krivan97}, good time
evolutions are achieved by writing the Teukolsky equation as a set
first order differential equations system. This and other issues
concerning the implementation of a fourth order algorithm are
discussed in the next chapter. Motivations for a fourth order
convergent algorithm are mainly, that it can reproduce the same
results of a second order convergent code with less resolution. This
makes the former to run faster than the later. Although in a fourth
order scheme, more intense computation is needed, there is some gain
in speed when equivalent resolutions, i.e. resolution that produce the
same error in the solution, are used. On the other hand, if equal
resolutions are used, a fourth order method will yield more accurate
solutions. The price we have to pay in this case is that the fourth
order method will take longer. This issues enter in consideration in
gravitational wave detection, because a large number of templates need
to be generated. The gain in accuracy can also be used in second
order perturbation theory, since higher order derivatives of the field
are needed to build up the effective source term \cite{manuela99}.
This may have important applications in the 'Lazarus approach'~\cite{Baker:2001sf}
and the radiation reaction problem of a particle orbiting a black hole.

Using the Kinnersley null tetrad:
\begin{equation}
\begin{array}{l}
\bfl^\mu = [(r^2 + a^2)/\Delta, 1, 0, a/\Delta] \\
\bfn^\mu = [r^2 + a^2, -\Delta, 0, a]/(2\Sigma) \\
\bfm^\mu = [ia\sin\theta,0,1,i\sin\theta]/[\sqrt{2}(r + ia\cos\theta)]
\end{array}
\end{equation}
where $\Sigma = r^2 + a^2\cos^2\theta$ and $\Delta = r^2 - 2Mr + a^2$
The Newman-Penrose equations written in Boyer-Lindquist coordinates lead to 
the Teukolsky equation \cite{teukolsky73}
\begin{eqnarray}
&&\lefteqn{\left[ \frac{(r^2+a^2)^2}{\Delta}-a^2 \sin^2 \theta \right] \partial_{tt} \Psi + \frac{4Mar}{\Delta}\partial_{t\phi}\Psi}  \nonumber \\
&&\mbox{}+2s\left[r-\frac{M(r^2-a^2)}{\Delta}+i\,a\cos\theta \right] \partial_t \Psi \nonumber \\
&&\mbox{}- \Delta^{-s} \partial_r(\Delta^{s+1} \partial_r \Psi) \nonumber \\
&&\mbox{}- \frac{1}{\sin\theta} \partial_\theta( \sin\theta \partial_\theta \Psi) - \left[\frac{1}{\sin^2 \theta}-\frac{a^2}{\Delta} \right] \partial_{\phi\phi} \Psi \nonumber \\
&&\mbox{}- 2s \left[ \frac{a(r-M)}{\Delta} + \frac{i \cos \theta}{\sin^2 \theta} \right] \partial_\phi \Psi + (s^2 \cot^2 \theta - s)\Psi \nonumber\\
&&\mbox{}= 4\pi\Sigma T.
\label{eq:teukeq}
\end{eqnarray}
The gravitational perturbations are recovered setting $s=\pm 2$, $s$
is a parameter called {\it spin weight}. Here the quantity of interest
is $\Psi = \rho^{-4} \psi_4^B$ with $\rho = -1/(r - ia\cos\theta)$ and
$s=-2$. For is $s=2$, $\Psi = \psi_0^B$.

In vacuum, it is well know that (\ref{eq:teukeq}) can be separated in
the frequency domain, by taking $\Psi=e^{-i\omega t} e^{im\phi}
S(\theta)R(r)$. When $s=0$, the functions $S(\theta)$ are the
spheroidal functions. When $a\omega=0$ these eigenfunctions are the
spin weighted spherical harmonics $_s Y^m_l(\theta,\phi) =
_s\!S^m_l(\theta) e^{im\phi}$. The general solution can be represented as
\begin{equation}
\Psi = \int d\omega \sum_{l,m} R(r)_sS^m_l(\theta)e^{im\phi}e^{i \omega t}.
\end{equation}
In most astrophysical applications, we are interested in computing
solutions that represent gravitational radiation at infinity. This
information is carried by $\psi_4^B$. This, together with the good
asymptotic behavior of the solutions are the motivations of choosing
the value $s=-2$ to perform the time evolutions. Knowing the value of
$\psi_4^B$ allows us to calculate the outgoing energy flux per unit
time as\cite{manuela99}
\begin{equation}
\frac{dE}{du} = \lim_{r\to\infty} \left[ \frac{r^2}{4\pi} \int_\Omega \,d\Omega\left| \int_{-\infty}^u d \tilde{u} \,\psi_4(\tilde{u}, r, \theta, \phi) \right|^2 \right]
\end{equation}
where $d\Omega = \sin \theta \,d\theta \,d\phi$ and $u=t-r$.

In the next section we explicitate the numerical techniques used to
solve the Teukolsky equation starting from a review of the second
order formalism developed in Ref.~\cite{krivan97}. We then derive the
forth order accurate in time and space algorithm, as a natural
generalization of the Lax-Wendorff algorithm. It is followed by a
stability analysis that find a Courant factor of 1.5 between the time
and spatial coordinate $r^*$ that ensures stable evolution.  We finish
Section II with a description of the radiative
boundary conditions imposed to the field in the two radial boundaries
and the angular boundary conditions imposed by the symmetry of the
solution. Section III deals with the applications of the fourth order
code developed. As initial data we consider an outgoing
Gaussian pulse located in the far region. The fourth order convergence
in both spatial variables in verified in Section III.B. We this we
compute the quasinormal modes and power-law tails and an ultimately
test of the code. Finally we discuss some of the consequences of
the computed decay powers on the light of some discussion in the 
literature.

%%%%%%%%%%%%%%%%%%%%%%%%%%%%%%%%%%%%%%%%%%%%%%%%%%%%%%%%%%%%%%%%%%%%%%%%
\section{Numerical Techniques}\label{Sec:techniques}

\subsection{Review}
%%%%%%%%%%%%%%%%%%%%%%%%%%%%%%%%%%%%%%%%%%%%%%%%%
In this work we will follow the approach used by Krivan
et. al. \cite{krivan97}. They integrate the Teukolsky equation in the
time domain by rewriting it as a set a first order partial
differential equations. The aim of their work is the evolution of
gravitational perturbations of the Kerr spacetime. This is done for
the case $s=-2$. This choice is due to the fact that for this
particular value of spin, the solutions are bounded for both, ingoing
and outgoing waves near the horizon ($r^* \to -\infty$) and at
infinity ($r^* \to \infty$). The asymptotic behavior of the solutions
of (\ref{eq:teukeq}), for a given spin weight $s$, representing
ingoing and outgoing waves is \cite{krivan97}, \cite{teukolsky73}
\begin{equation}
\lim_{r^*\to+\infty} |\Psi_s| \sim \left\{
\begin{array}{ll}
1/r^{2s+1} & \mbox{for outgoing} \\
1/r & \mbox{for ingoing}
\end{array} \right.
\label{eq:atinfty}
\end{equation}

\begin{equation}
\lim_{r^*\to-\infty} |\Psi_s| \sim \left\{
\begin{array}{ll}
1 & \mbox{for outgoing} \\
\Delta^{-s} & \mbox{for ingoing}
\end{array} \right.
\end{equation}

The procedure of solving the Teukolsky equation is initiated by
introducing the ansatz
\begin{equation}
\Psi(t,r^*,\theta,\widetilde{\phi}) = \sum_{m} r^3
e^{im\widetilde{\phi}} \Phi(t,r^*,\theta)
\label{eq:ansatz}
\end{equation}
where $r^*$ is the Kerr tortoise coordinate, defined as
\begin{equation}
dr^* = \frac{r^2+a^2}{\Delta}dr
\label{eq:tortoise1}
\end{equation}
So\footnote{Note that in the literature it is also used a different normalization for the logarithmic terms instead of $2M$ it appears $r_+$ and $r_-$ respectively.},
\begin{eqnarray}\label{eq:tortoise2}
r^*&=&r+\frac{r_+^2+a^2}{r_+-r_-}\ln\left|\frac{r-r_+}{2M}\right|
-\frac{r_-^2+a^2}{r_+-r_-}\ln\left|\frac{r-r_-}{2M}\right|,\nonumber\\
&&r_\pm=M\pm\sqrt{M^2-a^2}
\end{eqnarray}
and $\widetilde{\phi}$ is the Kerr azimuthal coordinate, defined as
\begin{equation}
d\widetilde{\phi} = d\phi + \frac{a}{\Delta}dr.
\label{eq:phitilde}
\end{equation}
So,
\begin{equation}\label{eq:phitilde2}
\widetilde{\phi}=\phi+\frac{a}{r_+-r_-}\ln\left|\frac{r-r_+}{r-r_-}\right|
\end{equation}

Now they introduce an auxiliary field $\Pi$ defined as
\begin{equation}
\Pi \equiv \partial_t \Phi + b \partial_{r^*} \Phi,
\label{eq:PIdef}
\end{equation}
where
\begin{equation}
b \equiv \frac{r^2+a^2}{\Sigma}
\label{eq:bdef}
\end{equation}
\begin{equation}
\Sigma^2 \equiv (r^2+a^2)^2-a^2\Delta \sin^2\theta.
\label{eq:sigmadef}
\end{equation}
This decomposes the second order differential Teukolsky equation
(\ref{eq:teukeq}) into a system of first order differential
equations. The resulting equation has the form
\begin{equation}
\partial_t {\bf u}+{\bf M}\partial_{r^*}{\bf u}+{\bf Lu}+{\bf Au} = 0
\label{eq:teukeq1}
\end{equation}
where ${\bf u}\equiv(\Phi_R,\Phi_I,\Pi_R,\Pi_I)^T$ is a column vector
whose components are the real and imaginary ($R,I$) parts of the
fields $\Phi$ and $\Pi$. The coefficients of the derivatives of
(\ref{eq:teukeq}) are rearranged as the elements of matrices ${\bf M}$
and ${\bf A}$, given by
\begin{equation}
{\bf M} \equiv \left[
\begin{array}{cccc}
b & 0 & 0 & 0 \\
0 & b & 0 & 0 \\
m_{31} & m_{32} & -b & 0 \\
-m_{32} & m_{31} & 0 & -b
\end{array} \right]
\end{equation}
\begin{equation}
{\bf A} \equiv \left[
\begin{array}{cccc}
0 & 0 & -1 & 0 \\
0 & 0 & 0 & -1 \\
a_{31} & a_{32} & a_{33} & a_{34} \\
-a_{32} & a_{31} & -a_{34} & a_{33}
\end{array} \right].
\end{equation}
The remaining matrix ${\bf L}$ contains the derivatives of the fields
with respect to the polar coordinate $\theta$, whose components are
\begin{equation}
{\bf L} \equiv \left[
\begin{array}{cccc}
0 & 0 & 0 & 0 \\
0 & 0 & 0 & 0 \\
l_{31} & 0 & 0 & 0 \\
0 & l_{31} & 0 & 0
\label{eq:l31}
\end{array} \right].
\end{equation}
The deduction of substitution (\ref{eq:PIdef}) and the value of each
element of the above matrices are given later in this chapter.

They proceed to solve the first-order system (\ref{eq:teukeq1}) using
the Lax-Wendroff method \cite{numrec}. For this, (\ref{eq:teukeq1}) is
written as
\begin{equation}
\partial_t {\bf u} + {\bf D} \partial_{r^*} {\bf u} = {\bf S},
\end{equation}
where ${\bf D}=\mbox{diag}(b,b,-b,-b)$ and ${\bf S}=-({\bf M}-{\bf
D})\partial_{r^*}{\bf u}-{\bf Lu}$.

To solve the above equation in Ref.~\cite{krivan97} it was used a grid
with 8000 points for $r^*$ and 32 points for $\theta$. The
computational domain was $-100M
\le r^*_i \le 500M$ and $0 \le \theta_j \le \pi$, with a Courant
condition of $\Dt \le \min(\Drs, 5\Dth)$.

Boundary conditions have been imposed as follows: $\Phi = \Pi = 0$ at the
horizon and outer boundary. Along the axis, $\Phi=0$ or
$\partial_\theta \Phi =0$ for $m$, (the azimuthal number) odd or
even, respectively.

With this settings Ref.~\cite{krivan97} code showed stability of the
order $1000M$ of evolution time. It was second order convergent for
times $<50M$, with a convergence rate higher than 1.3 for later times.

In the following sections we will concentrate in generalizing the
numerical algorithm to accomplish a fourth order convergent numerical
evolution, using the first-order formulation of the Teukolsky equation 
(\ref{eq:teukeq1}) as starting point.

%%%%%%%%%%%%%%%%%%%%%%%%%%%%%%%%%%%%%%%%%%%%%%%%%%%%
\subsection{Rewriting Teukolsky equation}
%%%%%%%%%%%%%%%%%%%%%%%%%%%%%%%%%%%%%%%%%%%%%%%%%%%%
\subsubsection{Separating $\phi$ dependence}
The Teukolsky equation is separable in the azimuthal variable for the
general case of $s\ne0$ and $a\ne0$. Furthermore, we will change the
normal Boyer-Lindquist coordinates $r$ and $\phi$ by $r^*$ and
$\widetilde{\phi}$, respectively. This has the advantage that $r$
approaches asymptotically to $r_+$ as $r^*$ goes to minus infinity, so
the inner boundary of the computational domain is approximated very
close to the horizon, but still outside the black hole.

The variable $\widetilde{\phi}$ is used to improve the behavior of the
coordinates near the horizon. This is a manifestation of the frame
dragging effect of a rotating black hole \cite{krivan96}.

Now we make use use of the ansatz (\ref{eq:ansatz}) used by Krivan
et. al., where they also include a factor $r^3$. This is done to
eliminate the increasing behavior of the solutions at infinity,
according to (\ref{eq:atinfty}). So we substitute (\ref{eq:ansatz})
into (\ref{eq:teukeq}), setting the source term $T=0$ to get vacuum
space solutions. After dropping a global $r^3 e^{im\widetilde{\phi}}$
factor we get
\begin{eqnarray}
\lefteqn{-\left[ \frac{(r^2+a^2)^2}{\Delta}-a^2 \sin^2 \theta \right] \partial_{tt}\Phi + \frac{2}{\Delta} \Big[ Ms(r^2-a^2) -rs\Delta \nonumber} \\
&&\mbox{}- ia(s\Delta \cos\theta + 2Mmr) \Big] \partial_t \Phi + \frac{(r^2+a^2)^2}{\Delta} \partial_{r^*r^*} \Phi \nonumber \\
&&\mbox{}+ \frac{1}{r\Delta} \left[ (8r^2+6a^2)\Delta - 2rs(r^2+a^2)(M-r)\right. \nonumber \\
&&\mbox{}\left. + 2iamr(r^2+a^2) \right] \partial_{r^*} \Phi 
+ \partial_{\theta\theta}\Phi + \cot\theta\,\partial_{\theta}\Phi \nonumber\\
&&\mbox{}+ \frac{1}{r^2 \Delta}\Big\{ 6\Delta^2 - r\Delta \Big[ 6M(s+1) -r(7s+6)\nonumber \\
&&\mbox{}+ r(s\cot\theta +m\csc\theta)^2 \Big]\nonumber \\
&&\mbox{} -2iamr \left[ 2rs(M-r) - 3\Delta \right] \Big\}=0
\label{eq:teukeqvac}
\end{eqnarray}
We can bring this equation into the form
\begin{eqnarray}
&&\partial_{tt}\Phi +C_{t}\partial_t\Phi + C_{r^*r^*}\partial_{r^*r^*}\Phi + C_{r^*} \partial_{r^*} \Phi \nonumber \\
&& + C_{\theta\theta}\partial_{\theta\theta}\Phi + C_\theta \partial_\theta \Phi + C_{\mbox{so}} \Phi = 0
\label{eq:teukeqc}
\end{eqnarray}
where the $C$'s represent the coefficients of the derivatives. We simple multiply by $-\Delta/\Sigma^2$. The results are:
\begin{eqnarray}
&&C_t = 2s\frac{M(a^2-r^2)+r\Delta}{\Sigma^2}+2ia\frac{2mMr+s\Delta\cos\theta}{\Sigma^2} \\
&&C_{r^*r^*} = -\frac{(r^2+a^2)^2}{\Sigma^2} \\
&&C_{r^*} = 2\frac{rs(M-r)(r^2+a^2)-(3a^2+4r^2)\Delta}{r\Sigma^2}\nonumber\\
&&\ \ \ \ \ \ \ \ -2iam\frac{r^2+a^2}{\Sigma^2}\\
&&C_{\theta\theta} = -\frac{\Delta}{\Sigma^2}\\
&&C_{\theta} = -\frac{\Delta}{\Sigma^2}\cot\theta\\
&&C_{\mbox{so}} = 2iam\frac{2rs(M-r)-3\Delta}{r\Sigma^2}+\nonumber\\
&&\Delta \frac{6Mr(s+1) -r^2(7s+6) -6\Delta + r^2(s\cot\theta +m\csc\theta)^2}{r^2\Sigma^2}\nonumber\\
\end{eqnarray}

With the Teukolsky equation written as (\ref{eq:teukeqc}), we may now
proceed to transform it into an equivalent first-order differential
equations system.

%%%%%%%%%%%%%%%%%%%%%%%%%%%%%%%%%%%%%%%%%%%%%%%%%%%%
\subsubsection{From second to first-order differential equations}
A common technique to numerically solve a second-order differential
equation is to rewrite it as a set of first-order differential
equations and apply the appropriate methods. For cases such as the
simple wave equation, getting such a first-order system is
straightforward. The equation reads
\begin{equation}
\partial_{tt} u = v_p^2 \partial_{xx} u,
\end{equation}
with $v_p$ the velocity of propagation and $u=u(t,x)$. It can be written as
\begin{eqnarray}
\partial_{tt} u &=& f \\
f &=& v_p^2 \partial_{xx} u.
\end{eqnarray}
Here $f$ is a function chosen in such a way that second derivatives
are eliminated. In this case $f=v_p\,\partial_{tx}v$ will do the job,
with $v=v(t,x)$, and we have
\begin{eqnarray}
\partial_{tt} u &=& v_p\,\partial_{tx}v \\
v_p\,\partial_{tx}v &=& v_p^2 \partial_{xx} u.
\end{eqnarray}
Now we factor out the time derivative in the first equation and the
spatial derivative in the second one and we get a system of
first-order equations
\begin{eqnarray}
\partial_{t}u &=& v_p\,\partial_{x}v  \\
\partial_{t}v &=& v_p\,\partial_{x}u 
\end{eqnarray}
or in matrix form
\begin{equation}
\left[ \begin{array}{c}
u \\
v
\end{array} \right]_t =
\left[ \begin{array}{cc}
0 & v_p \\
v_p & 0
\end{array} \right]
\left[ \begin{array}{c}
u \\
v
\end{array} \right]_x
\end{equation}
where the subscripts denote derivatives. Notice that it has the form
of the advection equation for a system of equations
\begin{equation}
\partial_t {\bf u} = {\bf V}_p\, \partial_x {\bf u}
\label{eq:advecvec}
\end{equation}
where ${\bf u}\equiv(u,v)^T$ is a column vector.

Trying to carry out the above procedure for the Teukolsky equation is
more complicated and probably will not yield results, because it
contains cross derivatives. Our approach here will be
different. Recall Eq.\ (\ref{eq:teukeqc}), in which we set the
coefficient of the second time derivative equal to one, dividing the
equation by precisely this coefficient (the subscripts in the $C$'s do
not mean differentiation with respect to the variables).

We want to write (\ref{eq:teukeqc}) as an advection equation with a
source term:
\begin{equation}
\left[ \begin{array}{c}
\Phi \\
\Pi
\end{array} \right]_t + 
\left[ \begin{array}{cc}
\beta_{11} & \beta_{12} \\
\beta_{21} & \beta_{22} 
\end{array} \right]
\left[ \begin{array}{c}
\Phi \\
\Pi
\end{array} \right]_{r^*} +
\left[ \begin{array}{cc}
\gamma_{11} & \gamma_{12} \\
\gamma_{21} & \gamma_{22} 
\end{array} \right]
\left[ \begin{array}{c}
\Phi \\
\Pi
\end{array} \right]= {\bf 0}.
\label{eq:advecteuk}
\end{equation}
where none of the $\beta$'s and $\gamma$'s depend on $t$. Expanding
the matrix products we get two equations
\begin{eqnarray}
\partial_t \Phi + \beta_{11}\partial_{r^*}\Phi + \beta_{12} \partial_{r^*}\Pi + \gamma_{11}\Phi + \gamma_{12}\Pi &=& 0,\nonumber\\
\label{eq:1st} \\
\partial_t \Pi + \beta_{21}\partial_{r^*}\Phi + \beta_{22} \partial_{r^*}\Pi + \gamma_{21}\Phi + \gamma_{22}\Pi &=& 0.\nonumber\\
\label{eq:2nd}
\end{eqnarray}
Now let $\beta_{12} = \gamma_{11} = 0$ and $\gamma_{12}=-1$. By doing
this we obtain an expression for $\Pi$ that depends only on the
derivatives of $\Phi$ with respect to $t$ and $r^*$ (and some function
$\beta_{11}$):
\begin{equation}
\Pi = \partial_t\Phi + \beta_{11} \partial_{r^*}\Phi,
\label{eq:Pidef2}
\end{equation}
which is similar to (\ref{eq:PIdef}), but $\beta_{11}$ has not yet
been specified. We will see that with this definition, $\Pi$ can
be easily substituted and eliminated in the second equation
(\ref{eq:2nd}). The derivatives of $\Pi$ are
\begin{eqnarray}
\partial_t\Pi &=& \partial_{tt}\Phi + \beta_{11}\partial_{t\,r^*}\Phi \label{eq:Pi_t}\\
\partial_{r^*}\Pi &=& \partial_{r^*t}\Phi + (\partial_{r^*}\beta_{11})( \partial_{r^*}\Phi) + \beta_{11} \partial_{r^*r^*}\Phi. \label{eq:Pi_r}
\end{eqnarray}
Substituting (\ref{eq:Pidef2}), (\ref{eq:Pi_t}) and (\ref{eq:Pi_r})
into (\ref{eq:2nd}) and rearranging terms we find
\begin{eqnarray}
\lefteqn{\partial_{tt}\Phi - \gamma_{22}\partial_t\Phi - (\beta_{11}+\beta_{22})\partial_{r^*t}\Phi + \beta_{22}\beta_{11}\partial_{r^*\,r^*}} \nonumber \\ 
&&- (\beta_{21} - \beta_{22} \partial_{r^*}\beta_{11} - \gamma_{22}\beta_{11})\partial_{r^*}\Phi - \gamma_{21}\Phi = 0.
\label{eq:betagamma}
\end{eqnarray}
To find the value of the $\beta$'s and $\gamma$'s we equate the
coefficients of the derivatives of $\Phi$ in (\ref{eq:betagamma}) with
those in (\ref{eq:teukeqc}). From the coefficient of
$\partial_{r^*t}\Phi$ we find that $\beta_{22} =
-\beta_{11}$. Combining this with the coefficient of
$\partial_{r^*r^*}\Phi$ we see that $\beta_{11}=\sqrt{-C_{r^*r^*}}$:
\begin{equation}
-\beta_{22}=\beta_{11} = \frac{r^2+a^2}{\Sigma} = b
\end{equation}
This is exactly the definition of $b$ in (\ref{eq:bdef}). It is easy to show that the remaining equations yield the following results:
\begin{eqnarray}
\gamma_{22} &=& C_t \\
\gamma_{21} &=& C_{\mbox{so}} + l_{31} \label{eq:l31_2}\\
\beta_{21} &=& C_{r^*} + b \,\partial_{r^*}b - C_t b. \label{eq:beta21}
\end{eqnarray}
Here $l_{31}$ is defined according to (\ref{eq:l31}) as the $\theta$-derivative operator:
\begin{equation}
l_{31} = C_{\theta\theta}\,\partial_{\theta\theta} + C_{\theta}\,\partial_{\theta},
\end{equation}
in the sense that it can be ``factored'' as $l_{31}\Phi$ and added to
$C_{\mbox{so}}\Phi$. Of course, an expression as (\ref{eq:l31_2}) is
not mathematically rigorous. It is rather a way to express that the
equation's angular dependence is going to be added to the source
term. Furthermore, in a numerical implementation we don't compute the
value of $\gamma_{21}$, but the value of $\gamma_{21}\Phi$.

%%%%%%%%%%%%%%%%%%%%%%%%%%%%%%%%%%%%%%%%%%%%%%%%
\subsubsection{Splitting real and imaginary parts}
Because Teukolsky equation involves complex coefficients it is
necessary to treat the real and imaginary parts of the solution. Let's
define four functions $\Phi_R, \Phi_I, \Pi_R$ and $\Pi_I$, such that
\begin{eqnarray}
\Phi &=& \Phi_R + i\Phi_I \\
\Pi  &=& \Pi_R + i\Pi_I
\end{eqnarray}
and substitute them into (\ref{eq:1st}) and (\ref{eq:2nd}). After
collecting real and imaginary parts and equating both to zero, we
obtain a set of four equations. As a shorthand to denote derivatives,
we use a dot for $\partial_t$ and a prime for $\partial_{r^*}$
\begin{equation}
\dot{\Phi}_R +b\Phi_R'-\Pi_R = 0
\end{equation}
\begin{equation}
\dot{\Phi}_I +b\Phi_I'-\Pi_I = 0
\end{equation}
\begin{equation}
\dot{\Pi}_R +\beta_{21}^R \Phi'_R -\beta_{21}^I\Phi'_I - b\Pi'_R +\gamma_{21}^R\Phi_R -\gamma_{21}^I\Phi_I + C_t^R\Pi_R - C_t^I\Pi_I=0
\end{equation}
\begin{equation}
\dot{\Pi}_I +\beta_{21}^I \Phi'_R +\beta_{21}^R\Phi'_I - b\Pi'_I +\gamma_{21}^I\Phi_R +\gamma_{21}^R\Phi_I + C_t^I\Pi_R + C_t^R\Pi_I=0.
\end{equation}
We can finally arrange these equations in matrix form as follows:
\begin{eqnarray}
&&
\left[ \begin{array}{c} 
\Phi_R \\
\Phi_I \\
\Pi_R \\
\Pi_I
\end{array} \right]_t +
\left[ \begin{array}{cccc}
b & 0 & 0 & 0 \\
0 & b & 0 & 0 \\
\beta_{21}^R & -\beta_{21}^I & -b & 0 \\
\beta_{21}^I & \beta_{21}^R & 0 & -b \\
\end{array} \right]
\left[ \begin{array}{c} 
\Phi_R \\
\Phi_I \\
\Pi_R \\
\Pi_I
\end{array} \right]_{r^*} +\nonumber\\
&&
\left[ \begin{array}{cccc}
0 & 0 & -1 & 0 \\
0 & 0 & 0 & -1 \\
\gamma_{21}^R & -\gamma_{21}^I & C_t^R & -C_t^I \\
\gamma_{21}^I & \gamma_{21}^R & C_t^I & C_t^R
\end{array} \right]
\left[ \begin{array}{c} 
\Phi_R \\
\Phi_I \\
\Pi_R \\
\Pi_I
\end{array} \right] = {\bf 0}.
\end{eqnarray}

Comparing this last equation with (\ref{eq:teukeq1}) and making use of
our definitions of $\beta_{21}$ and $\gamma_{21}$, equations
(\ref{eq:beta21}) and (\ref{eq:l31_2}), respectively; we see that our
derivation yields the same structure for Teukolsky equation as that
derived by Krivan et. al. \cite{krivan97}. It is worth to comment here
that although the structure of the equation is the same, the
coefficients shown above do not agree completely with those of
Ref.~\cite{krivan97}. The coefficients that do not agree are
$C_{\mbox{so}}$ and the real part of $C_{r^*}$, which correspond to
what Ref.~\cite{krivan97} calls $c_2, c_5$ and $c_6$. The coefficients
reported in \cite{krivan97} correspond exactly to the case in which
the ansatz for the solution is
\begin{equation}
\Psi(t,r^*,\theta,\widetilde{\phi}) = \sum_{m}
e^{im\widetilde{\phi}} \Phi(t,r^*,\theta).
\end{equation}
Notice that this equation does not contain the function $r^3$ used in
(\ref{eq:ansatz}). Another correction that needs to be done in order
to recover Teukolsky equation is to set $a_{31}=c_5$ and
$a_{34}=-c_3$, according to the definitions presented in
\cite{krivan97}.

Summarizing, the coefficients used in this work, using the ansatz
(\ref{eq:ansatz}) are
\begin{eqnarray}
\beta^R_{21} &=& 2\frac{rs(M-r)(r^2+a^2)-(3a^2+4r^2)\Delta}{r\Sigma^2} \nonumber \\
&& \mbox{}- 2bs\frac{M(a^2-r^2)+r\Delta}{\Sigma^2} +b \partial_{r^*} b\\
\beta^I_{21} &=& -2am\frac{r^2+a^2}{\Sigma^2} -2ba\frac{2mMr+s\Delta\cos\theta}{\Sigma^2} \\
\gamma^R_{21} &=& \frac{\Delta}{r^2\Sigma^2}\Big(6Mr(s+1) -r^2(7s+6) -6\Delta + \nonumber \\
&& \mbox{} r^2(s\cot\theta +m\csc\theta)^2\Big)\nonumber \\
&& \mbox{} -\frac{\Delta}{\Sigma^2}\partial_{\theta \theta} - \frac{\Delta}{\Sigma^2}\cot \theta \partial_{\theta} \\
\gamma^I_{21} &=& 2am\frac{2rs(M-r)-3\Delta}{r\Sigma^2} \\
C^R_t &=& 2s\frac{M(a^2-r^2)+r\Delta}{\Sigma^2} \\
C^I_t &=& 2a\frac{2mMr+s\Delta\cos\theta}{\Sigma^2}
\end{eqnarray}

%%%%%%%%%%%%%%%%%%%%%%%%%%%%%%%%%%%%%%%%%%%%%%%%
\subsection{Derivation of the 4th order algorithm}
%%%%%%%%%%%%%%%%%%%%%%%%%%%%%%%%%%%%%%%%%%%%%%%%
Let's assume that $u(t,x)$ is a continuously differentiable function
and that its derivatives in both $t$ and $x$ exist up to order four in
some given interval. This function satisfies a differential equation
of the form
\begin{equation}
u_t = v\, u_x
\label{eq:advection}
\end{equation}
where $v$ is a constant and the subscripts represent first derivatives
with respect to $t$ and $x$ respectively. The Taylor expansion in $t$
for $u(t,x)$ is
\begin{equation}
u(t+\Dt, x) = u + u_t \Dt  + u_{tt} \frac{\Dt^2}{2!}  + u_{ttt} \frac{\Dt^3}{3!} +  u_{tttt}\frac{\Dt^4}{4!} + \cdots.
\label{eq:taylorexp}
\end{equation}
Here $u$ and its derivatives are evaluated at some given point
$(t_0,x_0)$. Now we can use eq. (\ref{eq:advection}) to replace the
time derivatives by spatial derivatives, using the fact that
$u_{tt}=v^2 u_{xx}$, $u_{ttt}=v^3 u_{xxx}$ and so on. Thus we have
\begin{eqnarray}
u(t+\Dt,x) &=& u + v u_x\Dt + v^2 u_{xx}\frac{\Dt^2}{2!} + \nonumber \\ 
&&\mbox{} v^3u_{xxx} \frac{\Dt^3}{3!}  + v^4  u_{xxxx}\frac{\Dt^4}{4!} + \cdots.
\end{eqnarray}

If we truncate this series, taking terms up to $\Dt^2$ we will obtain
the Lax-Wendroff scheme, which is second order accurate in space and
time. Introducing the usual discrete notation $U^n_j$ for $u(t_n,x_j)$
and using a second order accurate approximation for $x$ derivatives we
have \cite{kincaid}
\begin{eqnarray}
U^{n+1}_j &=& U^n_j -\alpha \left( U^n_{j+1}-U^n_{j-1} \right) + \nonumber \\
&&\mbox{} \frac{1}{2}\alpha^2 \left(U^n_{j+1}-2U^n_j+U^n_{j-1} \right)
\end{eqnarray}
where $\alpha = v\Dt/\Dx$.

In order to have a fourth order accurate scheme, we truncate the
Taylor expansion of $u(t,x)$ including terms up to $\Dt^4$ and using
the differential equation (\ref{eq:advection}) to replace time
derivatives by space derivatives. Furthermore, we use a fourth order
finite difference scheme (see appendix \ref{ch:derivs}) to approximate
spatial derivatives. Proceeding in this way we have
\begin{eqnarray}
&& U^{n+1}_j = U^n_j +
\frac{\alpha}{12} \,\left( U^n_{j-2} - 
       8\,U^n_{j-1} + 8\,U^n_{j+1} - 
       U^n_{j+2} \right) - \nonumber \\
&&  \frac{{\alpha }^2}{24}\,
     \left( U^n_{j-2} - 16\,U^n_{j-1} + 
       30\,U^n_j - 16\,U^n_{j+1} + U^n_{j+2}
       \right) + \nonumber \\
&&  \frac{{\alpha }^3}{48}\,
     \left( U^n_{j-3} - 8\,U^n_{j-2} + 
       13\,U^n_{j-1} - 13\,U^n_{j+1} + 
       8\,U^n_{j+2} - U^n_{j+3} \right) \nonumber \\
&&   - \frac{{\alpha }^4}{144}\,
     \left( U^n_{j-3} - 12\,U^n_{j-2} + 
       39\,U^n_{j-1} - 56\,U^n_j + 
       39\,U^n_{j+1} \right. \nonumber \\
&&   \left. - 12\,U^n_{j+2} + 
       U^n_{j+3} \right). \nonumber \\
&&
\label{eq:44}
\end{eqnarray}
Notice how the dependence on the discretization parameters $\Dt$ and
$\Dx$ appears as powers of $\alpha$.

%%%%%%%%%%%%%%%%%%%%%%%%%%%%%%%
\subsection{Stability analysis}
%%%%%%%%%%%%%%%%%%%%%%%%%%%%%%%
We apply now a von Neumann stability analysis to the previous
scheme. This analysis is local, which means that we assume that the
coefficients of the finite difference equation vary slowly in space
and time such that they can be considered to be constant. In our case,
these coefficients do not depend on time. We say that the method is
stable if the scheme is stable for every constant value of the
coefficients in their range \cite{sewell}. The idea is to expand the
solution of the difference equation in its eigenmodes $e^{i k j \Dx}$,
where $k$ is a real wave number. The time dependence of these modes is
a succession of powers of some complex number $\xi(k)$, called {\it
amplification factor}. With this, we say that the difference equation
is stable if $|\xi(k)| \le 1$, for a given value of $k$. The
eigenmodes of the difference equation (\ref{eq:44}) can be written as
\cite{numrec}
\begin{equation}
U^n_j = \xi^n e^{ikj\Dx}
\label{eq:newmann}
\end{equation} 
where $\xi=\xi(k)$ is a complex quantity and $k$ is a real wave
number. Substituting (\ref{eq:newmann}) into (\ref{eq:44}) we get a
first grade polynomial in $\xi$. After some algebra we get
\begin{eqnarray}
\xi(k) &=& 
	1+ \frac{1}{3} \alpha^2( \cos k\Dx -7) \sin^2\frac{k\Dx}{2} - \nonumber \\ 
	&& \frac{2}{9} \alpha^4 (\cos k\Dx -4) \sin^4\frac{k\Dx}{2} + \nonumber \\ 
&&i \left[ \frac{1}{6}\alpha ( 8 \sin k\Dx - \sin 2k\Dx)+ \right. \nonumber \\
&&\left.\frac{1}{24} \alpha^3 (8 \sin 2k\Dx -13\sin k\Dx - \sin 3k\Dx) \right] \nonumber \\
&& 
\end{eqnarray}
or
\begin{eqnarray}
|\xi(k)|^2 &=&
1 + \frac{8}{9}\,{\alpha }^2\,
     \left(\cos k\Dx -5\right) \,
     {\sin^6 \frac{k\Dx}{2}} \nonumber \\
&& -\frac{4}{9}\,{\alpha }^4\,
     \left( 4\,\cos k\Dx -17 \right) \,
     {\sin^8 \frac{k\Dx}{2}} +\nonumber \\
&&  \frac{1}{54}\,{\alpha }^6\,
     \left( 133\,\cos k\Dx - 
       34\,\cos 2\,k\Dx + \right.\nonumber \\
&&    \left. 3\cos 3\,k\Dx-150 
       \right) \,{\sin^6 \frac{k\Dx}{2}} + \nonumber \\
&&   \frac{4}{81}\,{\alpha }^8\,
     {\left( \cos k\Dx -4\right) }^2\,
     {\sin^8 \frac{k\Dx}{2}}.
     \nonumber \\
&&
\label{eq:xi1}
\end{eqnarray}

\begin{figure}
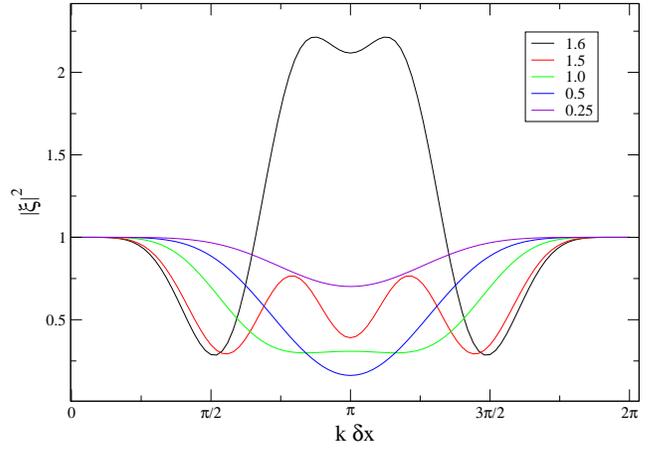

\centering
\includemygraphics[scale=0.35]{xi44.eps}
\caption{$|\xi|^2$ vs. $k\Dx$ for several values of $\alpha$.}
\label{fig:ampfact44}
\end{figure}

This is a periodic function with a period of $2\pi$. For the scheme to
be stable it has to satisfy the stability condition
\begin{equation}
|\xi| \le 1.
\end{equation} 
Although we can find analytic solutions for this equation, just to
present them here would occupy several pages. Rather than analytic
solutions, we are interested in some interval of values for $\alpha$,
such that our fourth order method is stable. The behavior of $|\xi|^2$
as a function of $k$ is shown in Fig.\ \ref{fig:ampfact44} for some
values of $\alpha$. With certainty, we can conclude that $\xi$ is less
than 1 for $\alpha \leq 1.5$. This implies that the Courant condition for
this case is
\begin{equation}
v\Dt \le 1.5 \Dx
\end{equation}

This value is $50\%$ greater than that required by the Lax-Wendroff
method and many others. This can be understood by expanding the
amplification factor $\xi(k)$ in series of $k\Dx$. In practice $\Dx$
must be small enough to correctly approximate the continuous
differential equation. This means that for modes corresponding to
small values of $k$ we can expand (\ref{eq:xi1}) in a power series of
$k\Dx$. That is
\begin{equation}
|\xi|^2 = 1 - \left( \frac{\alpha^2}{18}+\frac{\alpha^6}{72} \right) (k\Dx)^6 +
O((k\Dx)^8).
\end{equation}
It is interesting to compare this with the corresponding series
expansion of $|\xi|^2$ for the Lax and Lax-Wendroff methods. For the
Lax method we find \cite{numrec}
\begin{equation}
|\xi|^2 = 1-(1-\alpha^2)(k\Dx)^2 + \cdots
\end{equation}
and for the Lax-Wendroff method we have
\begin{equation}
|\xi|^2 = 1-\alpha^2(1-\alpha^2)\frac{(k\Dx)^4}{4}+\cdots.
\end{equation}
We can see that our fourth order method has a sixth order dependence
of $k\Dx$, which means that mode damping effects become relevant for
much higher values of $k$, making this method more accurate.  The
generalization of the scheme to two dimensions is illustrated in the
last subsection. In this case the situation is more
complicated because we are taking first and second derivatives in the
$\theta$ direction. We were not able to find an analytic Courant
factor that took into account the $\theta$ direction. From numerical
experiments we verify that the Courant condition used in
\cite{krivan97} was reliable in our case too. Thus, as a rule of
thumbs, we always kept $\Dt = \min( \Dr^*, 5 \Dth)$.

%%%%%%%%%%%%%%%%%%%%%%%%%%%%%%%%%%%%%%%%%%%%%%
\subsection{Boundary conditions}
%%%%%%%%%%%%%%%%%%%%%%%%%%%%%%%%%%%%%%%%%%%%%%
\subsubsection{Radial boundary conditions}
We use Sommerfeld boundary conditions at radial infinity and at the
event horizon. When one uses the tortoise coordinate $r^*$ the event
horizon is reached when $r^* \to -\infty$. In practice, it turns that
setting $r^*=-50M$ is a good approximation. For this value we have
$|r-2M| \approx 10^{-12}$. At the inner boundary, the condition is
that of an ingoing wave
\begin{equation}
\frac{\partial}{\partial t} \Phi(t,r^*,\theta) = \frac{\partial}{\partial r^*} \Phi(t,r^*,\theta).
\label{eq:innerBC}
\end{equation}
At the outer boundary, the appropriate condition is that of an outgoing wave
\begin{equation}
\frac{\partial}{\partial t} \Phi(t,r^*,\theta) = -\frac{\partial}{\partial r^*} \Phi(t,r^*,\theta).
\label{eq:outerBC}
\end{equation}
In order to make this conditions compatible with our fourth order
integration scheme, we take higher derivatives of (\ref{eq:innerBC})
and (\ref{eq:outerBC}). The idea is again, to substitute the time
derivatives of the Taylor expansion (\ref{eq:taylorexp}) by means of
the boundary conditions above. The results are summarized in table
\ref{ta:radialBC}.

\begin{table}
\centering
\begin{tabular}{c|c}
\hline \hline
Inner boundary $r^*=r^*_{min}$  &  Outer boundary $r^*=r^*_{max}$ \\
\hline
$\partial_t \Phi = \partial_{r^*} \Phi$  & $\partial_t \Phi = -\partial_{r^*} \Phi$  \\
$\partial^2_t \Phi = \partial^2_{r^*} \Phi$  & $\partial^2_t \Phi = \partial^2_{r^*} \Phi$ \\
$\partial^3_t \Phi = \partial^3_{r^*} \Phi$  & $\partial^3_t \Phi = -\partial^3_{r^*} \Phi$  \\
$\partial^4_t \Phi = \partial^4_{r^*} \Phi$  & $\partial^4_t \Phi = \partial^4_{r^*} \Phi$ \\
\hline \hline
\end{tabular}
\caption{Boundary conditions for the radial direction.}
\label{ta:radialBC}
\end{table}

The implementation of boundary conditions is straightforward when a
second order scheme is employed. This is due to the fact that we only
need up to second order spatial derivatives and its stencil demands
only 3 points\footnote{For centered finite differences, a second order
accurate formula needs two points to compute first derivatives and
three points for second derivatives.}. On the other hand, the case of
a fourth order accurate expression for spatial derivatives needs two
more points for the first derivative and seven points for the fourth
derivative, see appendix \ref{ch:derivs}. So, in this case, the way we
implemented the radial boundary conditions is to use off-centered
expressions to compute the spatial derivatives, when needed. Assuming
a computational grid of $N_r+1$ points in the $r^*$ direction,
labeling the points from 0 to $N_r$, table \ref{ta:offcent} shows the
points for which off-centered spatial derivatives are used.

\begin{table}
\centering
\begin{tabular}{c|c}
\hline \hline
Off-centered derivatives & used at points \\
\hline
$\partial_{r^*}, \partial^2_{r^*}$  & 1, $N_r-1$ \\
$\partial^3_{r^*}, \partial^4_{r^*}$  & 1, 2, $N_r-1$, $N_r-2$ \\
\hline \hline
\end{tabular}
\caption{Points for which off-centered derivatives in the $r^*$ direction are used (point labeling: $N_r+1$ points from 0 to $N_r$).}
\label{ta:offcent}
\end{table}

As to the auxiliary field $\Pi$, defined in (\ref{eq:PIdef}), its
boundary condition follows directly from its definition and from
(\ref{eq:innerBC}) and (\ref{eq:outerBC}). Once we know the boundary
values for $\Phi$, the value for $\Pi$ at the inner boundary is
\begin{eqnarray}
\Pi &=& \partial_t\Phi + b\,\partial_{r^*}\Phi \\
    &=& (b+1) \partial_{r^*} \Phi
\end{eqnarray}
where we have used (\ref{eq:innerBC}) and $b$ is given by
(\ref{eq:bdef}). In a similar way, the outer boundary condition is
\begin{equation}
\Pi = (b-1) \partial_{r^*} \Phi
\end{equation}

\subsubsection{Angular boundary conditions}
\begin{figure}
\centering
\includegraphics[scale=0.3]{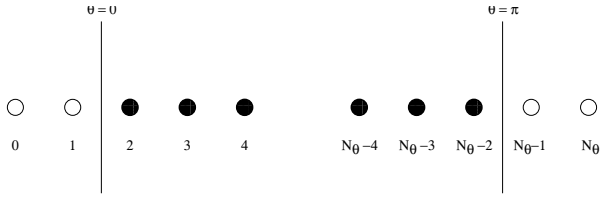}
\caption{White circles: ghost zones, black circles: normal grid points}
\label{fig:ghostzones}
\end{figure}

These are imposed along the rotation axis, i.e. at $\theta=0$ and
$\theta=\pi$. The boundary condition depends on the particular
azimuthal mode $m$ (see equation (\ref{eq:ansatz})) chosen for the
evolution, it can be stated as
\begin{eqnarray}
\Phi = 0 & \mbox{for} & m= \pm 1, \pm 3, \pm 5... \\
\partial_\theta \Phi=0 & \mbox{for} & m= 0, \pm 2, \pm 4...
\end{eqnarray}
This conditions come directly from the behavior of the solution in the
$\theta$ direction. At the same time it is precisely this behavior
what we use in order to implement the appropriate boundary
conditions. The solutions for which $m$ is even, have even parity
about both, $\theta=0$ and $\theta=\pi$. On the other hand, the
solutions with odd $m$, have odd parity about $\theta=0$ and
$\theta=\pi$, i.e.
\begin{eqnarray}
\left.
\begin{array}{rrr}
\Phi(t,r^*,\theta) &=& \Phi(t,r^*,-\theta) \\
\Phi(t,r^*,\pi+\theta) &=& \Phi(t,r^*,\pi-\theta)
\end{array} \right\} & \mbox{for} & m=0,\pm 2, ... \nonumber\\
\label{eq:BCth0} \\
\left.
\begin{array}{rrr}
\Phi(t,r^*,\theta) &=&-\Phi(t,r^*,-\theta) \\
\Phi(t,r^*,\pi+\theta)&=&-\Phi(t,r^*,\pi-\theta)
\end{array} \right\} & \mbox{for} & m=\pm 1,\pm 3, ...\nonumber\\ 
\label{eq:BCthpi}
\end{eqnarray}
To take advantage of this property we need to use a staggered grid in
the $\theta$ direction. By doing this we also avoid the inherent
difficulties of evaluating expressions in which $\cot\theta$ is
present, (like the last term of Teukolsky equation (\ref{eq:teukeq}))
since this function is not finite neither at $\theta=0$ nor
$\theta=\pi$. In a staggered grid, the values for $\theta=0$ and
$\theta=\pi$ are always located exactly between two grid points. The
points to the left (right) of $\theta=0$ ($\theta=\pi$) are considered
as ``ghost zones'', because are used just to implement the boundary
conditions. In our fourth order method, we need four ghost points. Two
before the first point immediately after $\theta=0$ and two more after
the point immediately before $\theta=\pi$, as shown in Fig.\ 
\ref{fig:ghostzones}. Our grid has $N_\theta+3$ points in the $\theta$
direction, the first two and last two points are ghost zones. In this
way we can always use a centered formula to compute the derivatives in
the $\theta$ direction. Notice that if we are using a second order
accurate approximation, we only need one ghost point at each end of the
grid. The values of the ghost zones are updated according to
(\ref{eq:BCth0}) and (\ref{eq:BCthpi}).

These properties of the solutions are a direct consequence of the
spherical harmonics behavior (when the spin parameter $s=0$) and the
spin weighted spherical harmonics (when $s=-2$).

%%%%%%%%%%%%%%%%%%%%%%%%%%%%%%%%%%%%%%%%%%%%%%%%%%%%%%
\subsection{Notes on implementation}
%%%%%%%%%%%%%%%%%%%%%%%%%%%%%%%%%%%%%%%%%%%%%%%%%%%%%%
We mention here some implementation details of equation
(\ref{eq:teukeq1}), which for convenience will be written as\footnote{We
use lower case bold face to denote vectors and upper case bold face to
denote matrices.}
\begin{equation}
\partial_t \bfu = -\bfM \partial_{r^*} \bfu - ( \bfL + \bfA ) \bfu.
\label{eq:finalteuk}
\end{equation}
As stated above, the main idea has been always to substitute the time
derivatives in the Taylor expansion (\ref{eq:taylorexp}) by spatial
derivatives using our differential equation
(\ref{eq:finalteuk}). Calling $\bfB = -\bfM$ and $\bfS = -(\bfL + \bfA
)$, all time derivatives needed are:
\begin{eqnarray}
\partial_t \bfu &=& \bfB \partial_{r^*} \bfu + \bfS \bfu \\
\partial^2_t \bfu &=& \bfB \partial_{r^*} ( \partial_t \bfu ) + \bfS (\partial_t \bfu )\\
\partial^3_t \bfu &=& \bfB \partial_{r^*} ( \partial^2_t \bfu ) + \bfS (\partial^2_t \bfu )\\
\partial^4_t \bfu &=& \bfB \partial_{r^*} ( \partial^3_t \bfu ) + \bfS (\partial^3_t \bfu ).
\end{eqnarray}
Here we have used the fact that none of the coefficient of the
Teukolsky equation are time dependent and that partial derivatives
commute.

The four time derivatives above could be computed using just the
finite differences formula for the first derivative in $r^*$. Once we
know $\partial_t \bfu$ we can {\it numerically} substitute this result
to get the second time derivative $\partial^2_t \bfu$ and so on. The
problem with this procedure is that because of the exclusive use of
the first $r^*$ derivative formula; at the end, we will not have a
fourth $r^*$ derivative with the accuracy shown in appendix
\ref{ch:derivs}. It still will be fourth order accurate but using the
first derivative formula four times will propagate more error than
that of the fourth order accurate finite differences formula. The same
holds for the other derivatives. The approach we took was to compute
all time derivatives directly from the coefficients of the evolution
equation, and the value of the fields at every time step. Of course,
this implies much more larger expressions to compute time derivatives,
because now we have to {\it algebraically} substitute one time
derivative into the other. Carrying out such a substitutions we find:
\begin{eqnarray}
\partial_t \bfu &=&  \bfB \,\bfu' + \bfS \,\bfu   \\
\partial^2_t \bfu &=& \bfB\,\left[ \bfB'\,\bfu' + (\bfS\,\bfu)' + \bfB\,\bfu'' \right] + \bfS \,\partial_t \bfu\\
\partial^3_t \bf\bfu &=& \bfB\,
   \left\{ (\bfS \,\partial_t \bfu)' + {\bfB'}^2\,\bfu' + 
     \bfB'\,\left[ (\bfS\,\bfu)' + 3\,\bfB\,\bfu'' \right] \right. \nonumber \\
&&  \left. \mbox{} +   \bfB\,\left[ \bfu'\,\bfB'' + (\bfS\,\bfu)'' + 
        \bfB\,\bfu^{(3)} \right]  \right\} + \bfS \, \partial^2_t \bfu \\
\partial^4_t \bfu &=& \bfB\,
   \left\{ (\bfS \, \partial^2_t \bfu)' + {\bfB'}^3\,\bfu' + 
     {\bfB'}^2\,\left( (\bfS\,\bfu)' + 7\,\bfB\,\bfu'' \right) +\right. \nonumber \\
&&  \mbox{} \bfB'\,\left[ (\bfS \, \partial_t \bfu)' + 
        \bfB\,\left( 4\,\bfu'\,\bfB'' + 3\,(\bfS\,\bfu)'' + 
           6\,\bfB\,\bfu^{(3)} \right)  \right]  \nonumber \\
&& \left. \mbox{} + \bfB\,\left[ (\bfS\,\bfu)'\,\bfB'' + 
        (\bfS \, \partial_t \bfu)'' + 
        \bfB\,\left( 4\,\bfB''\,\bfu'' \right.\right.\right.\nonumber \\
&& \left.\left.\left. + \bfu'\,\bfB^{(3)} + 
           (\bfS\,\bfu)^{(3)} + \bfB\,\bfu^{(4)} \right)
        \right]  \right\} \nonumber \\
&& \mbox{} + \bfS \, \partial^3_t \bfu.
\end{eqnarray}
To clarify the equations, we have used ``primes'' to denote
differentiation with respect to $r^*$.

There is one final issue, worth mentioning here. Notice that there are
still some products in which the time derivatives of $\bfu$ and $\bfu$
itself appear explicitly. All of these products involve multiplication
with $\bfS$. Neither the time derivatives nor $\bfu$ are algebraically
substituted in these products because $\bfS=-(\bfL + \bfA )$ and
$\bfL$ contains the $\theta$ derivatives operator. Instead of
expanding further derivatives, we chose to numerically substitute the
time derivatives of $\bfu$ and $\bfu$ itself into these products. We
use the term ``numerically substitute'' in the sense that each time
derivative is calculated and stored in the memory of the computer,
further computation makes use of the stored values.

Such a procedure gives good results as shown in the next section.

%%%%%%%%%%%%%%%%%%%%%%%%%%%%%%%%%%%%%%%%%%%%%%%%%%%%%%%%%%%%%%%%%%%%%%%%
\section{Results}\label{Sec:results}

\subsection{Initial data}
%%%%%%%%%%%%%%%%%%%%%%%%%%%%%%%%%%%%%%%%%%%%%%
In all runs, initial data with compact support was used. The function
used was a Gaussian bell centered at $r^*=75M$, in the $r^*$ direction
and some $l,m$ mode dependence in the $\theta$ direction. Thus, for
$t=0$ we have
\begin{equation}
\Phi(0,r^*,\theta) = e^{-(r^*-75)^2/100} \Theta_{lm}(\theta)
\label{eq:ID}
\end{equation}
where $\Theta_{lm}(\theta)$ represents the $\theta$ dependence as
spherical harmonics $Y_l^m(\theta, \phi)$ or the spin weighted
spherical harmonics $_sY_l^m(\theta,\phi)$; for $s=0$ or $s=-2$,
respectively. Table \ref{ta:sphericalh} shows the $\theta$ dependence
of the first four spherical harmonics and spin weighted spherical
harmonics.

\begin{table}
\centering
\begin{tabular}{c|c|c}
\hline \hline
$l$-mode& $Y_l^0$  &  $_{-2}Y_l^0$ \\
\hline
$l=0$ & constant & -- \\
$l=1$ & $\cos\theta$ & -- \\
$l=2$ & $3\cos^2\theta-1$ & $\sin^2\theta$ \\
$l=3$ & $5\cos^3\theta-3\cos\theta$ & $\cos\theta\sin^2\theta$ \\
$l=4$ & $35\cos^4\theta-30\cos^2\theta+3$ & $(5+7\cos2\theta)\sin^2\theta$ \\ 
\hline \hline
\end{tabular}
\caption{$\theta$ dependence of $Y_l^0$ and $_{-2}Y_l^0$ without the normalization constant.}
\label{ta:sphericalh}
\end{table}

%%%%%%%%%%%%%%%%%%%%%%%%%%%%%%%%%%%%%%%%%%%%%%%%%%%%
\subsection{Fourth order convergence}
%%%%%%%%%%%%%%%%%%%%%%%%%%%%%%%%%%%%%%%%%%%%%%%%%%%%

Convergence was tested in both $r^*$ and $\theta$ directions. The
method to assess convergence was to compare three runs for the same
initial data but different resolutions. If we want to measure
convergence in the $r^*$ direction, we keep $\theta$ resolution fixed;
while we vary $r^*$ resolution. The same holds in the case of
assessing $\theta$ convergence. The way of varying resolutions is such
that they keep the same ratio. If we call this resolutions {\it fine,
medium} and {\it coarse}; they satisfy:
\begin{equation}
\frac{\Drs_{coarse}}{\Drs_{medium}}= \frac{\Drs_{medium}}{\Drs_{fine}}=\rho_{r^*}
\end{equation}
where $\rho_{r^*}$ is some positive number. In practice, this ratio
was taken to be 1.5 or 2. Each time the resolution is increased, the
numerical solution must converge to the true solution. The numerical
solution will have an error of the order of $(\Dr^*)^4$, then we can
say that
\begin{eqnarray}
\Psi_{coarse} &=& \Psi_{true} + k (\Dr^*)^4 \\
\Psi_{medium} &=& \Psi_{true} + k (\Drs/\rho_{r^*})^4 \\
\Psi_{fine} &=& \Psi_{true} + k (\Drs/\rho_{r^*}^2)^4 
\end{eqnarray}
where $k$ is a constant. From these relations it is easy to verify that
\begin{equation}
\rho_{r^*}^4 (\Psi_{fine} - \Psi_{medium}) = \Psi_{medium} - \Psi_{coarse}
\label{eq:diffres}
\end{equation}
This means that when plotted together, $\rho_{r^*}^4 (\Psi_{fine} -
\Psi_{medium})$ and $\Psi_{medium} - \Psi_{coarse}$ must lie on top of
each other and that will indicate that our numerical scheme is fourth
order convergent. The same applies in the case of $\theta$.

\begin{figure}
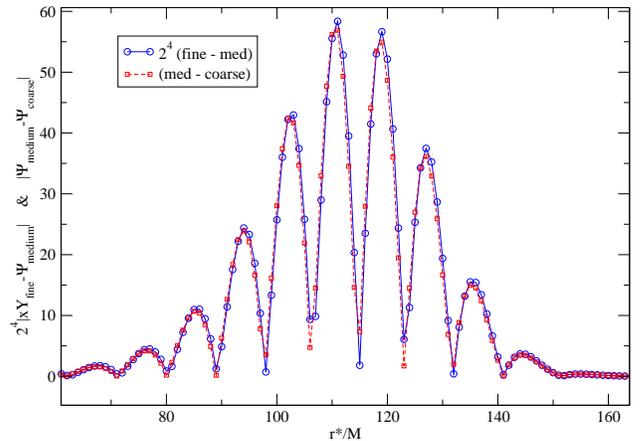

\centering
\includemygraphics[scale=0.35]{convr1.eps}
\caption{Convergence test in $r^*$ at $\theta=\pi/2$ and $t=200M$.}
\label{fig:convr1}
\vskip .7cm
\end{figure}

\begin{figure}
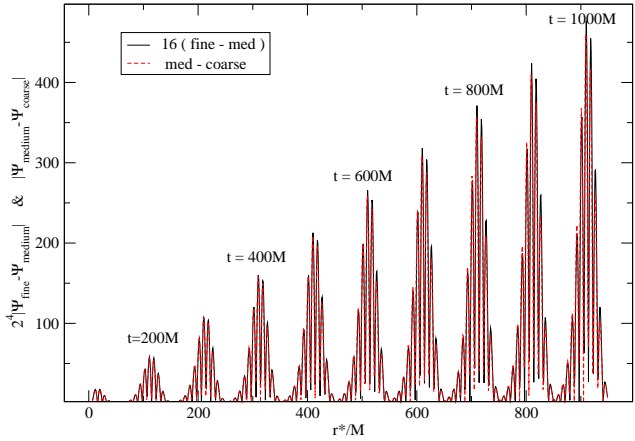

\centering
\includemygraphics[scale=0.35]{convr4.eps}
\caption{Convergence test in $r^*$ at $\theta=\pi/2$ at time intervals of $100M$.}
\label{fig:convr4}
\end{figure}

Starting with the $r^*$ direction, Figs.\ 
\ref{fig:convr1}-\ref{fig:convr4} shows fourth order convergence. The
simulation parameters are (in all runs the black hole mass is taken as
$M=1$ and the Courant factor is 0.5):

\begin{center}
\begin{tabular}{ll}
\hline
domain: & $-50M \le r^* \le 950M$ \\
grid size $(N_{r^*} \times N_\theta)$: & $1000 \times 8$ \\
resolution: & $\Drs=\{1,0.5,0.25\}M$, $\Dth=\pi/8$ \\
physical parameters: & $a=0,l=2,m=0,s=-2$ \\
initial data: & ingoing Gaussian pulse\\
\hline
\end{tabular}
\end{center}

Figure \ref{fig:convr1} shows the differences (\ref{eq:diffres}) in
absolute value. It corresponds to the instant $t=200M$. By this time
the initial pulse has bounced off the potential barrier and has reached
a maximum amplitude of $\sim 5\times 10^3$. A rough estimation
indicates an error of 0.08\% at the highest amplitud. Fig.\ 
\ref{fig:convr2} shows the same phase than the previous one but at
$t=1000M$. The error has increased now to $\sim 0.6\%$. A carefull
examination of Fig.\ \ref{fig:convr2} shows that the difference
between the two lines is more notorious than in Fig.\ 
\ref{fig:convr1}. That means that the convergence ratio is less than
4, but it is still consistent within a ratio of 3.95.
Fig.\ \ref{fig:convr4} shows the same phase of the previous two
graphs at different time steps. It is very clear that the relative
error increases linearly with time.

\begin{figure}
\centering
\includemygraphics[scale=0.35]{convr2.eps}
\caption{Convergence test in $r^*$ at $\theta=\pi/2$ and $t=1000M$.}
\vskip .7cm
\label{fig:convr2}
\end{figure} 

\begin{figure}
\centering
%stagth.eps is replace by stagth2.eps because of bounding box
\includemygraphics[scale=0.42]{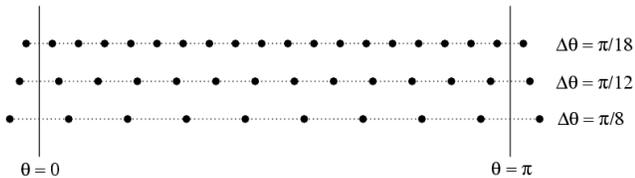}
\caption{Distribution of the grid points $\theta_k$ for different resolutions.}
\label{fig:stagth}
\end{figure}

As for the convergence in the $\theta$ direction, a minor problem needs to
be solved before computing the differences of the three numerical
solutions. The problem is that the implementation of a staggered grid
in $\theta$ causes that the discrete set of values $\theta_k$ were
completely different when resolution is changed. This fact is
illustrated in Fig.\ \ref{fig:stagth}. Therefore, in order to assess
convergence, we used a sixth order Lagrange polynomial to interpolate
the solution obtained with the medium and finer resolutions at the
values $\theta_k$ of the coarse one.  Having done that, Fig.\ 
\ref{fig:convth1} shows fourth order convergence for a fixed value of
$r^*=20M$. The simulation parameters are still the same but this time
$\Drs=0.25M$ and $\Dth=\{\pi/16, \pi/24, \pi/36\}$, i.e. the ratio
$\rho_\theta = 1.5$. The amplitude decreases as the wave passes
by. The graph shows different snapshot at intervals of
$200M$. Convergence is lost when round-off error is reached.

\begin{figure}
\centering
\includemygraphics[scale=0.35]{convth1_2.eps}
\caption{Convergence test in $\theta$ at $r^*=20M$ at time intervals of $200M$.
The continuous lines represent the fine minus the medium resolution (times 
$1.5^4$) and the dashed one the medium minus the coarse one.}
\label{fig:convth1}
\end{figure}

\begin{figure}
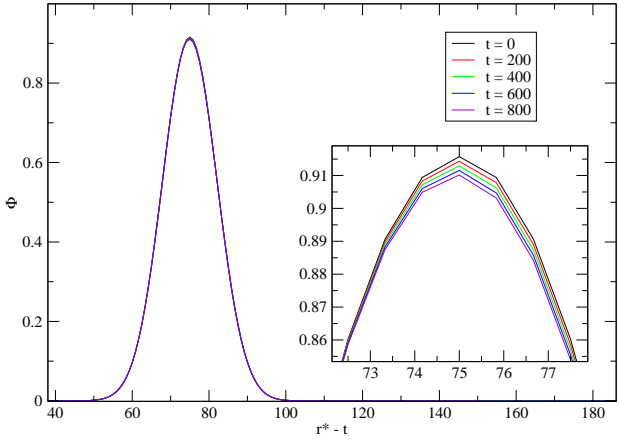

\centering
\includemygraphics[scale=0.35]{source_test.eps}
\caption{Snapshots of an outgoing Gaussian pulse. Numerical dispersion
effects appear at late times (space and time are in units of M).}
\label{fig:source_test}
\end{figure}

Other test performed to check the validity of the numerical solutions
was evolving the initial data of a known analytic function. This
function does not need to be a solution of the Teukolsky equation, it
could be any smooth function in $r^*$ and $\theta$, provided that the
corresponding source term is added to the evolution equations. The
idea is the following: let's call $\mathcal{T}$ the ``Teukolsky
operator'' so that $\mathcal{T}(\Phi(t,r^*,\theta))=0$ is the
Teukolsky equation (\ref{eq:teukeqvac}). If we choose an arbitrary
smooth function $\widetilde{\Phi}(t,r^*,\theta)$, the result of applying the
operator $\mathcal{T}$ will be
\begin{equation}
\mathcal{T}(\widetilde{\Phi}(t,r^*,\theta)) = f(t,r^*,\theta).
\end{equation}
If we add the source term $f(t,r^*,\theta)$ to the evolution equations
and give the initial data as $\widetilde{\Phi}(0,r^*,\theta)$ and
$\partial_t\widetilde{\Phi}(0,r^*,\theta)$ our code should reproduce
the function $\widetilde{\Phi}$. In Fig.\ \ref{fig:source_test} we
show the result of such a test. We set the function $\widetilde{\Phi}$
as a Gaussian pulse (in $r^*$ and $\theta$), traveling in the
increasing direction of $r^*$. Our code reproduce the analytic
function with high accuracy. There is some damping in the amplitude of
the pulse due to numerical dispersion; however, this effect appears at
very late times.

%%%%%%%%%%%%%%%%%%%%%%%%%%%%%%%%%%%%%%%%%%%
\subsection{Power-law tails}
%%%%%%%%%%%%%%%%%%%%%%%%%%%%%%%%%%%%%%%%%%%
The main application of this work is to accurately
compute the power-law falloff in
the gravitational perturbations evolution. In the following
results the observation point is located at $r^*=20$M and
$\theta=\pi/2$. The highest resolution used was $\Dr^*=0.125$M,
$\Dth=\pi/48$ and the lowest one was $\Dr^*=1$M,
$\Dth=\pi/8$. Variation of the resolution in those intervals was done
in order to verify convergence, although the lowest resolution was not
enough in cases where the $\theta$ profile presents several
oscillations.  Computationally, the determination of power-law tails
is a challenging problem, because the amplitude of the wave decays
exponentially during the quasinormal ringing and as an inverse power
of time during the tail phase. This means that we are working with
very small numbers that eventually reach round-off error, due to the
finite precision of the computer processor. This posses some
difficulties when we try to determine the exponent of the
power-law. Recall that at very late times, for a finite value of $r^*$
(timelike infinity), the amplitude of the field goes as
\begin{equation}
\Phi \propto t^{-(2l+3)}.
\label{eq:pricefallof}
\end{equation}
In principle, finding this exponent should not be a problem since the
field is proportional to a power of the time $t$. Thus, a simple power
fitting of the form $\Phi = At^{-\mu}$ (where $A$ and $\mu$ are
constants) would be just enough. The problem with this procedure is
that (\ref{eq:pricefallof}) is the very last stage in the evolution of
the perturbation. In practice, we are not able to evolve the
perturbations for such a long time, due to the finiteness of
computational resources\footnote{In our case this finiteness is
precisely RAM memory. This is because the outer boundary condition is
not perfect and after some finite time, part of the initial pulse is
reflected back. To delay the arrival of this reflection the outer
boundary must be far away from the observation point, which means a
large computational domain; in other words: more RAM memory.}. So we
analyze the field $\Phi$ just from the moment the tail phase begins
until the moment the solution reaches round-off error. During this
period, the field falloff is governed also by powers of time smaller
than $-(2l+3)$ \cite{scheel04}:
\begin{equation}
\Phi \propto t^{-\mu} + O(t^{-(\mu+1)})
\end{equation}
which means that the exponent of $t$ reaches $-(2l+3)$ in an
asymptotic way. Taking this fact in consideration, we compute the
``local power index'' \cite{burko97} defined as $\mu_N=-t \,\partial_t
\Phi / \Phi$, and use a linear fit (least-squares) such that
\begin{equation}
\mu_N = \mu + \frac{B_1}{t} + \frac{B_2}{t^2},
\label{eq:linear}
\end{equation}
where the $B$'s are constants.

\begin{figure}
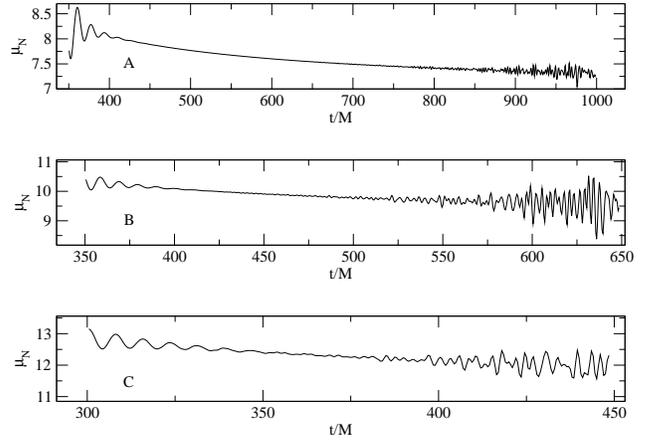

\centering
\includemygraphics[scale=0.35]{localpower_diffl.eps}
\caption{Local power index $\mu_N$ for $m=0, a=0$. Part A, B and C correspond to $l$ values of 2, 3 and 4; respectively, with their corresponding power-law tails of 7, 9 and 11.}
\label{fig:localpower}
\vspace{0.7cm}
\end{figure}

The least-squares fit yields good results when the local power index
$\mu_N$ (that is computed taking numerical derivatives of $\Phi$) is
taken in a large interval over which its oscillations are small. This
behavior of $\mu_N$ is illustrated in Fig.\ \ref{fig:localpower}. For
times longer than those shown in this plot, the oscillation of $\mu_N$
becomes larger and the meaning of the local power index is lost.

\begin{figure}
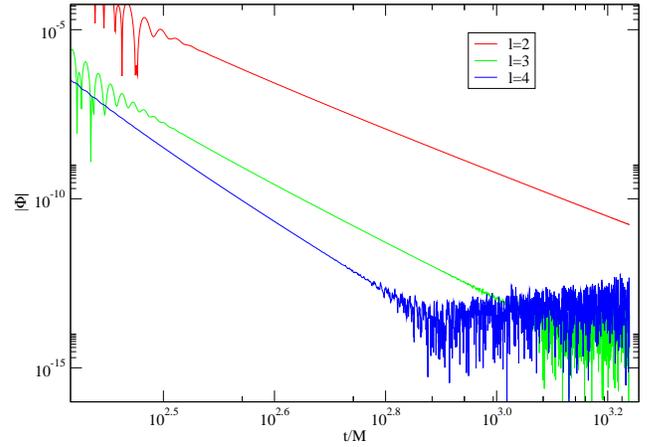

\centering
\includemygraphics[scale=0.35]{tails_diffl_m0_a0.eps}
\caption{Power-law tails for $m=0$, $a=0$ and different values of $l$. The evolution time is 1600M with a resolution $\Dr^*=0.25$M and $\Dth=0.098$.}
\label{fig:tails_diffl_m0_a0}
\end{figure}

We notice that the larger the value of $l$ the shorter the interval of
validity for $\mu_N$. This is due to the fact that for larger values
of $l$, the power-law exponent is bigger (in absolute value), making
the field to decay very fast reaching round-off error
earlier. Fig.\ \ref{fig:tails_diffl_m0_a0} shows this behavior for
$m=0$, $a=0$ and different values of $l$. A closer examination of
Figs.\ \ref{fig:localpower} and \ref{fig:tails_diffl_m0_a0} reveals
that the oscillations in $\mu_N$ start some time before round-off
error appears. The origin of such behavior is attributed to the
accumulated numerical error that increases as the evolution
progresses. This oscillations are magnified in Fig.\ 
\ref{fig:localpower} due to the numerical time derivatives of the
field $\Phi$.

\begin{table}
\centering
\begin{tabular}{cccccc}
\hline \hline
$l$ & $m$ & predicted $\mu$ & $\mu$  $O(t^{-1})$ & $\mu$ $O(t^{-1}, t^{-2})$ & $\mu$ $O(t^{-2})$  \\
\hline
2 & 0 & 7 & 6.866 $\pm$ 0.004 & 7.00 $\pm$ 0.03 & 7.18 $\pm$ 0.01 \\ 
2 & 1 &   & 6.87 $\pm$ 0.01 & 7.01 $\pm$ 0.03   & 7.177 $\pm$ 0.003 \\ 
2 & 2 &   & 6.87 $\pm$ 0.01 & 7.04 $\pm$ 0.06   & 7.180 $\pm$ 0.005 \\ 
\hline
3 & 0 & 9 & 8.64 $\pm$ 0.04 & 8.9 $\pm$ 0.3     & 9.24 $\pm$ 0.02\\
3 & 1 &   & 8.62 $\pm$ 0.04 & 8.9 $\pm$ 0.2     & 9.22 $\pm$ 0.02\\ 
3 & 2 &   & 8.62 $\pm$ 0.02 & 8.9 $\pm$ 0.1     & 9.23 $\pm$ 0.01\\
3 & 3 &   & 8.65 $\pm$ 0.04 & 9.1 $\pm$ 0.2     & 9.23 $\pm$ 0.02\\
\hline
4 & 0 & 11& 10.1 $\pm$ 0.7 & 11.3 $\pm$ 0.7     & 11.23 $\pm$ 0.04 \\
%4 & 1 &   & NA & NA &  NA \\
%4 & 2 &   & NA & NA &  NA \\
4 & 3 &   & 9.9 $\pm$ 0.1 & 10.7 $\pm$ 0.7      & 11.06 $\pm$ 0.04 \\
4 & 4 &   &10.2 $\pm$ 0.1 & 11.5 $\pm$ 0.5      & 11.31 $\pm$ 0.03 \\
\hline \hline
\end{tabular}
\caption{Numerically computed power-law tails, using a least-squares fit to $\mu_N$.}
\label{ta:tails1}
\end{table}

\begin{table}
\centering
\begin{tabular}{ccccccccc}
\hline \hline
    &     && $O(t^{-1})$  && \multicolumn{2}{c}{$O(t^{-1},t^{-2})$} && $O(t^{-2})$ \\
$l$ & $m$ && $B_1$ && $B_1$ & $B_2$ && $B_2$ \\
\hline
2 & 0 &&  460 && 220 & 76278 &&  144986\\
2 & 1 &&  440 && 234 & 71298 &&  151267\\
2 & 2 &&  439 && 198 & 82955 &&  150505\\
\hline
3 & 0 &&  577 && 314 & 61318 &&  134175\\
3 & 1 &&  583 && 346 & 55274 &&  135607\\
3 & 2 &&  580 && 306 & 64049 &&  134996\\
3 & 3 &&  569 && 173 & 92460 &&  132637\\
\hline
4 & 0 &&  798 && 502 & 91111 &&  145278\\
4 & 3 &&  885 && 253 & 112895 && 158033\\
4 & 4 &&  783 && 136 & 164384 && 140118\\
\hline \hline
\end{tabular}
\caption{Coefficients computed using the linear fitting.}
\label{ta:coeff_fit}
\end{table}

Table \ref{ta:tails1} shows the results a linear fit of the form
(\ref{eq:linear}) for the case $a=0$ and different values of $l$ and
$m$. The fourth column shows the value of the exponent $\mu$ using a
fitting curve of the form $\mu_N = \mu + B_1/t$. The fifth column
corresponds to a fitting curve $\mu_N = \mu + B_1/t + B_2/t^2$ and the
sixth column takes in consideration only a quadratic correction in
$t$, i.e. $\mu_N = \mu + B_2/t^2$. The uncertainty is the
statistically computed error for the parameter $\mu$ in the
fitting. This quantity is greater in the case of corrections
$O(t^{-1}, t^{-2})$ because there are three constants to be
adjusted. The exponents are in agreement with the expected value
$2l+3$. The agreement is better for $l=2$ than for $l=4$ and for the
$O(t^{-1},t^{-2})$ correction than for the $O(t^{-1})$ and $O(t^{-2})$
one. In two cases ($l=4$, $m=1,2$), the duration of the tail was not
enough to determine the exponent. The interval chosen to make the
curve fitting was the largest one that starts after the quasinormal
ringing and ends before the amplitude of the $\mu_N$ oscillations got
too high in a way that it could bias the result. In table
\ref{ta:coeff_fit} we show the values of the coefficients $B_1$ and
$B_2$ for the three different fitting functions. It is very
interesting to notice that the weight of the term $t^{-2}$ is greater
than that of $t^{-1}$ by a factor of at least 200. This result
supports, to some extend, the model proposed by Poisson
\cite{poisson02} for the radiative falloff of a scalar field in a
stationary, asymptotically flat and weakly curved spacetime. He shows
that the first correction to the power-law tail is of order
$t^{-2}$. Our observation point is located at $r^*=20M$ and it is not
far enough from the black hole to say that it is in the asymptotically
flat region.

\begin{figure}
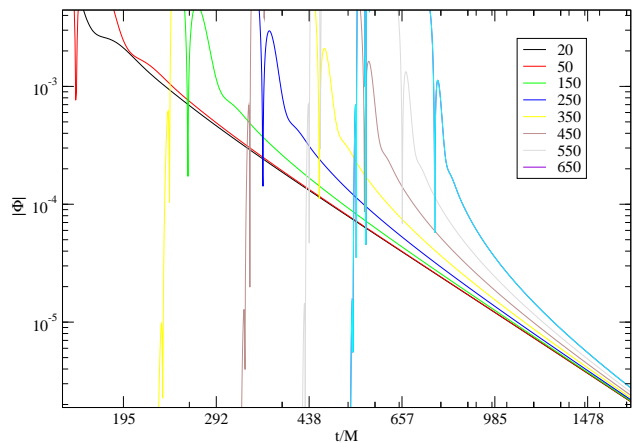

\vspace{0.7cm}
\centering
\includemygraphics[scale=0.35]{scalar_tails.eps}
\caption{Power law tails, scalar case $l=0, m=0$. Observer's position is in M units.}
\label{fig:scalar_tails}
\end{figure}

To find out the behavior in the asymptotically flat region we did
similar evolutions for the scalar case $s=0, l=0$ and $m=0$. Tables
\ref{ta:scalar_out} and \ref{ta:scalar_sym} show the power law tail
computed at different observation points in the equatorial plane. In
the first one the initial data was an outgoing Gaussian pulse centered
at $r^*=100$M. In the second one, initial data has the same initial
shape but the pulse has zero velocity, i.e. it is time symmetric. We
see that the power law tail is 4 instead of 3 for time symmetric
initial data. The fact that the fitting to a function of order
$t^{-2}$ yields tails closer to the predicted value for distant
observer supports Poisson's formula. Figure \ref{fig:scalar_tails}
show the exponential fall off at different distances from the black
hole as a function of time. All them approach asymptotically to the
theoretically known value.

\begin{table}
\begin{tabular}{c|c|c|c}
\hline \hline 
observ  & $O(t^{-1})$  &  $O(t^{-2})$  &  $O(t^{-1}, t^{-2})$ \\
\hline
20  & 2.969 & 2.951 & 3.001 \\  
50  & 2.965 & 2.951 & 3.001 \\ 
150 & 2.922 & 2.951 & 3.005 \\
250 & 2.825 & 2.967 & 3.025 \\
350 & 2.649 & 2.928 & 3.095 \\
450 & 2.474 & 2.807 & 3.190 \\
550 & 2.595 & 2.939 & 3.104 \\
650 & 2.377 & 2.906 & 3.234 \\
\hline \hline
\end{tabular}
\caption{Power law tails, scalar case $l=0$. Initial data: outgoing pulse. (Observer position in M units)}
\label{ta:scalar_out}
\end{table}

\begin{table}
\begin{tabular}{c|c|c|c}
\hline \hline 
observ  & $O(t^{-1})$  &  $O(t^{-2})$  &  $O(t^{-1}, t^{-2})$ \\
\hline
20  & 3.959 & 4.005 & 4.001 \\  
50  & 3.950 & 4.005 & 4.002 \\ 
150 & 3.859 & 3.999 & 4.019 \\
250 & 3.652 & 3.981 & 4.079 \\
350 & 3.601 & 3.979 & 4.084 \\
450 & 3.296 & 3.941 & 4.219 \\
550 & 3.419 & 3.961 & 4.143 \\
650 & 3.128 & 3.921 & 4.291 \\
\hline \hline
\end{tabular}
\caption{Power law tails, scalar case $l=0$. Initial data: zero velocity pulse. (Observer position in M units)}
\label{ta:scalar_sym}
\end{table}

\begin{figure}
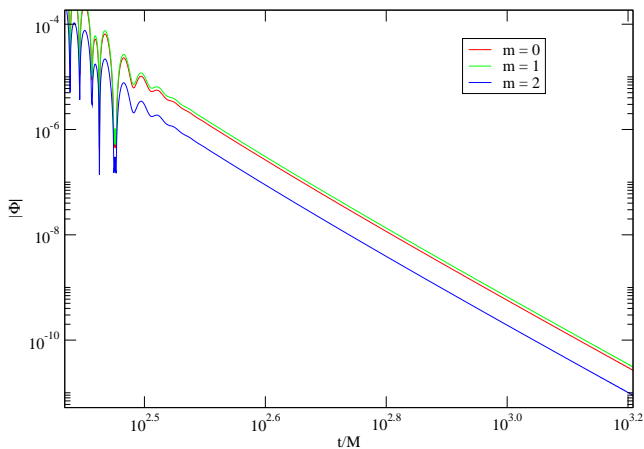

\centering
\vspace{.7cm}
\includemygraphics[scale=0.35]{l2_diffm_a0.eps}
\caption{Power-law tail $l=2$, $m=0,1,2$, $a=0$.}
\label{fig:l2_diffm_a0}
\end{figure}

\begin{figure}
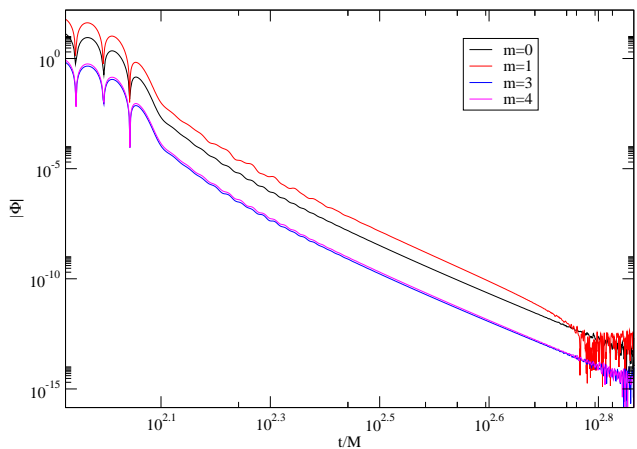

\centering
\vspace{0.7cm}
\includemygraphics[scale=0.35]{l4_diffm_a0.eps}
\caption{Power-law tail $l=4$, $m=0,1,3,4$, $a=0$.}
\label{fig:l4_diffm_a0}
\end{figure}

Figure \ref{fig:l2_diffm_a0} shows the evolution for the $l=2$
multipole for different values of $m$. In these cases there is no
presence of round-off error because it has the slowest decay rate,
$\Phi \propto t^{-7}$. In Fig.\ \ref{fig:l3_diffm_a0} we see the same
situation as above but with $l=3$. The power-law has a behavior $\Phi
\propto t^{-9}$. Round-off error appears at $\Phi \sim
10^{-12}$. Finally in Fig.\ \ref{fig:l4_diffm_a0} round-off error
appears approximately at the same value of $\Phi$ as in the previous
case. We notice that the quasinormal ringing is practically
inexistent. In all the runs we use outgoing initial data as prescribed
in (\ref{eq:ID}).

\begin{figure}
\centering
\vspace{0.7cm}
\includemygraphics[scale=0.35]{l3_diffm_a0.eps}
\caption{Power-law tail $l=3$, $m=0,1,2,3$, $a=0$.}
\label{fig:l3_diffm_a0}
\end{figure}

\begin{figure}
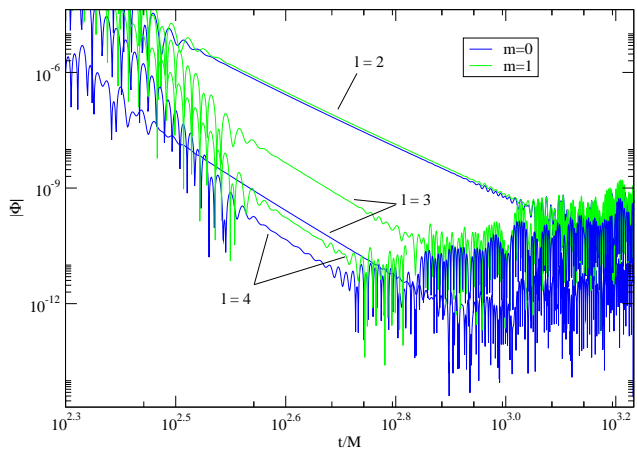

\centering
\vspace{0.7cm}
\includemygraphics[scale=0.35]{tails_kerr.eps}
\caption{Power-law tails for $a=0.5$.}
\label{fig:tails_kerr}
\end{figure}

So far we have been considering power-law tails for the case of a
non-rotating black hole ($a=0$). Fig.\ \ref{fig:tails_kerr} shows the
power-law tails for the case in which $a=0.5$. We can see that for
these values of $l$ the duration of the tail is too small. Doing a
nonlinear fitting for the case $l=2$, we found a tail of
$-7.0011$. For the other cases, this very short tail phase is not
enough to tell with certainty the power-law exponent; besides, the
tail has still some small oscillation in that time interval. To verify
the effects of a Kerr spacetime in the evolution of the gravitational
perturbations, the frequencies of the quasinormal ringing are useful;
as shown in the next section.

\begin{figure}
\centering
\includemygraphics[scale=0.35]{l3_m0_diffa.eps}
\caption{Power-law tail $l=3$, $m=0$, $a=0,0.5,0.9$.}
\label{fig:l3_m0_diffa}
\vspace{0.7cm}
\end{figure}

\begin{figure}
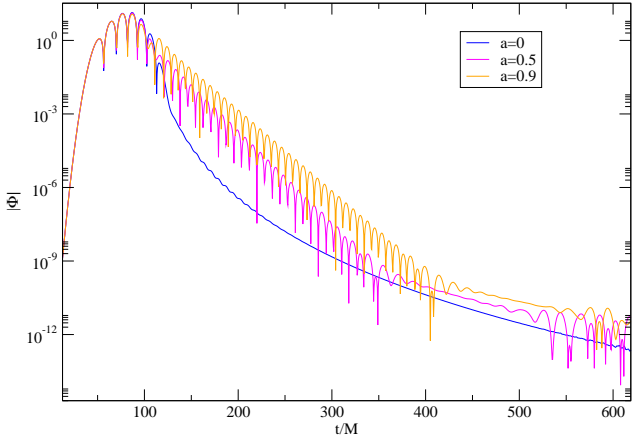

\centering
\includemygraphics[scale=0.35]{l4_m0_diffa.eps}
\caption{Power-law tail $l=4$, $m=0$, $a=0,0.5,0.9$.}
\label{fig:l4_m0_diffa}
\end{figure}

%%%%%%%%%%%%%%%%%%%%%%%%%%%%%%%%%%%%%%%%%%%%%
\subsection{Quasinormal modes}
We compute the quasinormal modes frequency for the cases shown in
Fig.\ \ref{fig:tails_kerr}, which correspond to the Kerr spacetime
(table \ref{ta:qnm_kerr}). The case $l=2$ agrees with the known
frequencies \cite{seidel90, kokkotas91} with an error less than
$1\%$. The cases $l=3$ and $l=4$ have frequencies similar to the $l=2$
multipole. We also compute these frequencies for the case of a
Schwarzschild spacetime (table \ref{ta:qnm_schwa}). In this case, the
frequency values agree with the predicted ones
\cite{leaver85,giammatteo03, froman92} within a 0.1 to $1.4\%$
error. In the cases for $l=4$, the numerical evolution was not able to
render quasinormal ringing. For $a \ne 0$, angular mode conversion to
the lowest allowed multipole is present. The case $l=3$, $m=0$
presents an irregular oscillation and the real part of the frequency
and cannot be trusted. That is the reason why it is not shown here.

\begin{table}
\centering
\begin{tabular}{cccc}
\hline \hline 
$l$ & $m$ & computed $\sigma$ M & predicted $\sigma$M \\
\hline
2 & 0 &  $0.384 + 0.0875i$ & $0.3833 + 0.08707i$\\
2 & 1 &  $0.341 + 0.0805i$ & $0.4206 + 0.08617i$\\
\hline
3 & 0 &  NA $ \mbox{}  + 0.0800i$ & $0.61212 + 0.09077i$\\
3 & 1 &  $0.339 + 0.0803i$ & $0.65060 + 0.0900i$\\
\hline
4 & 0 &  $0.382 + 0.0860i$ & NA\\
4 & 1 &  $0.341 + 0.0797i$ & NA\\
\hline \hline
\end{tabular}
\caption{Quasinormal mode frequencies for $a=0.5$ (NA = not available).}
\label{ta:qnm_kerr}
\end{table}

\begin{table}
\centering
\begin{tabular}{cccc}
\hline \hline
$l$ & $m$ & $\sigma$ M & predicted $\sigma$M \\
\hline
2 & 0 & $0.373 + 0.0875i$  & $0.3736715 + 0.0889625i$ \\
2 & 1 & $0.375 + 0.0869i$  & $0.3736715 + 0.0889625i$\\
2 & 2 & $0.376 + 0.0877i$  & $0.3736715 + 0.0889625i$\\
\hline
3 & 0 & $0.601 + 0.0903i$  & $0.5994435 + 0.092703i$\\
3 & 1 & $0.600 + 0.0902i$  & $0.5994435 + 0.092703i$\\
3 & 2 & $0.605 + 0.0901i$  & $0.5994435 + 0.092703i$\\
3 & 3 & $0.598 + 0.0933i$  & $0.5994435 + 0.092703i$\\
\hline \hline
\end{tabular}
\caption{Quasinormal mode frequencies for $a=0$.}
\label{ta:qnm_schwa}
\end{table}

Finally, Figs.\ \ref{fig:l3_m0_diffa} and \ref{fig:l4_m0_diffa} show
the evolution for $m=0$ of $l=3$ and $l=4$ respectively, for different
values of $a$. We can see roughly that the power-law tail is the same
for each case and that multipole conversion is present.

%%%%%%%%%%%%%%%%%%%%%%%%%%%%%%%%%%%%%%%%%%%%%%%%%%
\subsection{Fourth order versus second order}
%%%%%%%%%%%%%%%%%%%%%%%%%%%%%%%%%%%%%%%%%%%%%%%%%%
We could say that the advantage of using a fourth order convergent
code is that we can achieve the same results of the second order one
with less resolution. In other words, the same degree of accuracy can
be obtained with both approaches in the same computational domain but
the second order one will need more points. Quantitatively, we can
compare the error in the solution for both cases. This error is the
difference between the true solution and the numerical solution for a
given resolution. If we denote this quantity by $e_h$, where $h$ is
the grid spacing then we have:
\begin{equation}
e_h = k_n h^n
\label{eq:error}
\end{equation}
where $k$ is a constant and $n$ is the order of accuracy. If we equate the errors for $n=2$ and $n=4$, the relationship between resolutions is
\begin{equation}
h_2 = \frac{k_4}{k_2} h_4^2.
\end{equation}
If in the fourth order method, $h_4=0.1$ then the equivalent
resolution in the second order one is approximately 10 times bigger! 
i.e. $h_2=0.01$. This implies that in a one dimensional problem the
number of points is also 10 times bigger. If a two dimensional problem
is considered then the second order grid should contain 100 times more
points than that of the fourth order method, to get the same error in
the solution.

A feature of the finite differences methods is that according to
(\ref{eq:error}), if we increase resolution by some factor $c$ the error
is reduced by a factor $c^n$. Thus each time we double resolution, the
error decreases by a factor of 4 in the second order method and by 16
in the fourth order method.

The above considerations put the fourth order method in a better
position than the second order one but, as we said in the
introduction, the price we have to pay is running time. Given a
resolution $h$, the fourth order method will find a more accurate
solution than the second order one. The time that the fourth order
method will take will be longer because there are much more
calculations to be done. In order to determine if the gain in a
smaller grid is greater than the loss in running time, we did some
numerical experiments. The running time $t_{run}$ is given
approximately by
\begin{equation}
t_{run} \sim \frac{\Dth_0}{\Dth} \left( \frac{\Dr^*_0}{\Dr^*} \right)^2 t_0,
\end{equation}
where $t_0$ is the running time at resolutions $\Dr^*_0$ and
$\Dth_0$. Fixing $\Dr^*_0=1$ and $\Dth_0=\pi/8$, $t_0 = 16.3$ min in
the case of the fourth order method; and $t_0=3.5$ min for the second
order one (we used a Pentium 4 CPU 2.4GHz). These times correspond to
the gravitational case $l=2$, $m=0$ in an interval $-50<r^*<950$M
being the simulation time 1600 M. In this simulation, the fourth order
method gave very good results whereas the second order one becomes
unstable around $t=300$M. In order to obtain the same result
(power-law tail) than the fourth order one, it was necessary the
increase the resolution four times in both directions. This implies
that now, the running time for the second order method is $t_{run}
\sim 230$ min. This time is 14 times larger than that corresponding to
the fourth order method. So we definitely have a gain in speed, when
the errors in the numerical solutions (for both methods) are kept
equal.

As to the RAM memory, table \ref{ta:ram} gives information about the
amount of memory used in function of the grid size ($N_r \times
N_\theta$). These values depend on the coding details of the
algorithm. In this kind of problem, computer memory is not a crucial
factor as it is the speed.

\begin{table}
\centering
\begin{tabular}{ccccc}
\hline \hline
 & \multicolumn{2}{c}{RAM memory [Mb]} & \multicolumn{2}{c}{running time [hrs]} \\
$N_r \times N_\theta$ & 2nd order & 4th order & 2nd order* & 4th order \\
\hline
$1000 \times 8$  & 19  & 22  & 0.06 & 0.25 \\
$2000 \times 16$ & 64  & 71  & 0.48 & 2\\
$4000 \times 32$ & 236  & 249  & 3.73 & 16\\
\hline \hline
\end{tabular}
\caption{RAM memory used grid different grid sizes for both second and fourth order methods. *These times are for equal grid sizes. For equivalent resolutions, the running time for the 2nd order method is approximately 14 times larger than that of the 4th order method.}
\label{ta:ram}
\end{table}

%%%%%%%%%%%%%%%%%%%%%%%%%%%%%%%%%%%%%%%%%%%%%%%%%%%%%%%%%%%%%%%%%%%%%%%%%%

\section{Conclusions}

The main goal of this research has been to implement a stable fourth
order accurate method to numerically integrate the Teukolsky equation
in the time domain. In order to verify and evaluate the efficacy of
our fourth order method, we have reproduced the main known results,
i.e. power-law tails and quasinormal ringing. Power-law tails is a
subject in which there are still unsolved questions concerning the
late time behavior of the perturbations and its dependence on the
coordinates and the initial data~\cite{poisson02}. We addressed some
of these issues here and others are left for future research.
Among the formers, we studied the case of an initial pulse with
an initial configuration such that $\Phi_\ell|^{t=0}\not=0$ and
$\partial_t\Phi_\ell|^{t=0}=0$. We have been able in this case that
the late time behavior agrees well with the predicted~\cite{Karkowski:2003et}
decay $\sim 1/t^{2\ell+4}$. We have also confirmed numerically
that for an observer located far away from the hole $(r_{Obs}>20M)$
we see the predicted~\cite{poisson02} correction to the power law
as $\sim A/t^{2\ell+3}+B/t^{2\ell+5}$.

The fourth order method implemented in this work has shown to be
convergent and stable in all runs we did. The ansatz proposed by
Krivan et. al. \cite{krivan97} effectively removes the growing in time
of the field, a feature that is expected from the asymptotic behavior
of the solutions. When carrying out our calculation, we notice that
the coefficients of the Teukolsky equation reported in \cite{krivan97}
do not correspond to those we got when the ansatz proposed is
used. Instead, they correspond to an ansatz without the $r^3$ factor
(see eq. (\ref{eq:ansatz})). We also verify that the formulation of
Teukolsky equation as a system of first order differential equations
is a powerful technique because it allows to express the first time
derivative in function of the spatial derivatives. The resulting
system of equations has the form of the advection equation. That was
the key idea that allowed us to implement a fourth order method to
solve the equation: expand the solution in power series of time
keeping terms up to fourth order and then use the differential
equation to substitute the time derivatives by spatial
derivatives. Using this procedure, a higher order method could be
developed in a straightforward manner. The stability analysis provided
a limiting value of the Courant of 1.5 below which we can perform stable
evolutions.

The time evolutions carried out with the fourth order method yielded
accurate results even when relatively low resolutions were used. The
lowest resolution was $\Dr^*=1$, $\Dth=\pi/8$ for the $l=2$, $m=0$
case. The highest resolution used was $\Dr^*=0.125$ and $\Dth=\pi/48$,
for the $l=4$ multipole. In finite difference methods the resolution
is chosen in such a way that the details in the profiles of the
functions involved in the calculations can be accurately
approximated. For higher values of $l$, the $\theta$-dependence has
more oscillations in its domain therefore more points are needed to
find a reliable solution. The more oscillations or narrow peaks a
function has the more resolution we need. This comes from the fact
that those functions have derivatives whose values oscillate rapidly
and higher derivatives vary even faster.  If the resolution is not
good enough the effect is a ``numerical mode mixing'', that has
nothing to do with the physical model, but with the numerical aspects
of the implementation. This mode mixing acts like if we were evolving
the waves in a Kerr background, where physics tells us that angular
mode mixing is expected. This unwanted effect cannot be easily
detected when $a \ne 0$ in the simulations. Therefore it is absolutely
necessary to verify that this effect is not present when we evolve in
the Schwarzschild background, where the physical mode mixing does not
happen. This may explain some discrepancy on the computed power-law
exponent decay appeared in the literature.

For $a \ne 0$ table~\ref{ta:qnm_kerr} shows clearly that for $\ell=3,4,..$
the mode mixing acts bringing down the quasinormal frequencies close
to that of the $\ell=2$. The same effect is observed in the tails
power, although it is more difficult to prove (See Fig.~\ref{fig:tails_kerr}).

Boundary conditions were not an issue of concern in this research. The
radial inner boundary was not a problem because the field decays
exponentially near that region. The radial outer boundary always
reflects part of the wave. The immediate solution is to push this
boundary far away such that this reflection does not interfere in the
region of interest. A refinement of the boundary conditions will be
needed when a large computational domain can not be used in favor of
higher resolutions.

We have a reliable computational tool to explore several interesting
problems concerning first order perturbations. We expect to shed some
light on problems like the late time behavior of gravitational and
scalar fields in a Kerr background. This is a problem that has been
studied both analytically and numerically in recent years and there
are still questions to be answered. Further research could include the
problem of the orbiting particle around a black hole and the close
limit approximation in the problem of two colliding black holes
\cite{lousto01}.

%%%%%%%%%%%%%%%%%%%%%%%%%%%%%%%%%%%%%%%%%%%%%%%%%%%%%%%%%%%%%%%%%%%%%%%%%%

\acknowledgments 
We are very grateful to E. Poisson for useful discussions on
power-law tails, and to K. Kokkotas for making available unpublished
values of the Quasinormal frequencies for rotating black holes.  We also
gratefully acknowledge the support of the NASA Center for
Gravitational Wave Astronomy at The University of Texas at Brownsville
(NAG5-13396) and also to NSF for financial support from grants
PHY-0140326 and PHY-0354867.

\appendix

\widetext
\section{Fourth order accurate derivatives}\label{ch:derivs}

This are the formulas to compute fourth order accurate derivatives
using finite differences. Those which are centered have always less
truncation error than the corresponding off-centered ones. In all of
them, $h$ is the size of the step and $x \le \xi \le x+h$. We neglect
term of powers higher than four.

\begin{itemize}
\item First derivative
\begin{itemize}
\item centered:
\begin{equation}
u'(x) = \frac{{u_{j-2}} - 8\,{u_{j-1}} +  8\,{u_{j+1}} - {u_{j+2}}}{12\,h} - \frac{1}{30}h^4 u^{(5)}(\xi)
\end{equation}
\item off-centered (1 point):
\begin{equation}
u'(x) = \frac{-3\,{u_{j-1}} - 10\,{u_j} +  18\,{u_{j+1}} - 6\,{u_{j+2}} +  {u_{j+3}}}{12\,h} + \frac{1}{20}h^4u^{(5)}(\xi)
\end{equation}
\item off-centered (2 points):
\begin{equation}
u'(x)= \frac{-25\,{u_j} + 48\,{u_{j+1}} - 36\,{u_{j+2}} + 16\,{u_{j+3}} - 3\,{u_{j+4}}}{12\,h} - \frac{1}{5}h^4 u^{(5)}(\xi)
\end{equation}
\end{itemize}

\item Second derivative
\begin{itemize}
\item centered:
\begin{equation}
u''(x) = \frac{-{u_{j-2}} + 16\,{u_{j-1}} -  30\,{u_j} + 16\,{u_{j+1}} - {u_{j+2}}}{12\,h^2} -\frac{1}{90}h^4u^{(6)}(\xi)
\end{equation}

\item off-centered (1 point):
\begin{eqnarray}
u''(x)&=&\frac{10\,{u_{j-1}} - 15\,{u_j} - 4\,{u_{j+1}} + 14\,{u_{j+2}} - 6\,{u_{j+3}} + {u_{j+4}}}{12\,h^2} \nonumber \\
&& \mbox{}- \frac{13}{180}h^4 u^{(6)}(\xi)
\end{eqnarray}

\item off-centered (2 points):
\begin{eqnarray}
u''(x) &=& \frac{45\,{u_j} - 154\,{u_{j+1}} + 214\,{u_{j+2}} - 156\,{u_{j+3}} + 61\,{u_{j+4}} - 10\,{u_{j+5}}}{12\,h^2} \nonumber \\
&& \mbox{} -\frac{137}{180}h^4 u^{(6)}(\xi)
\end{eqnarray}
\end{itemize}

\item Third derivative
\begin{itemize}
\item centered:
\begin{eqnarray}
u'''(x) &=& \frac{{u_{j-3}} - 8\,{u_{j-2}} + 13\,{u_{j-1}} - 13\,{u_{j+1}} + 8\,{u_{j+2}} - {u_{j+3}}}{8\,h^3} \nonumber \\
&& \mbox{} - \frac{7}{120}h^4 u^{(7)}(\xi)
\end{eqnarray}

\item off-centered (1 point):
\begin{eqnarray}
u'''(x)&=& \frac{-{u_{j-2}} - 8\,{u_{j-1}} + 35\,{u_j} - 48\,{u_{j+1}} + 29\,{u_{j+2}} - 8\,{u_{j+3}} + {u_{j+4}}}{8\,h^3} \nonumber \\
&& \mbox{}+\frac{1}{15}h^4u^{(7)}(\xi)
\end{eqnarray}

\item off-centered (2 points):
\begin{eqnarray}
u'''(x)&=& \frac{-15\,{u_{j-1}} + 56\,{u_j} - 83\,{u_{j+1}} + 64\,{u_{j+2}} - 29\,{u_{j+3}} + 8\,{u_{j+4}} - {u_{j+5} }}{8\,h^3} \nonumber \\
&& \mbox{} - \frac{7}{120}h^4 u^{(7)}(\xi)
\end{eqnarray}

\item off-centered (3 points):
\begin{eqnarray}
u'''(x)&=& \frac{-49\,{u_j} + 232\,{u_{j+1}} - 461\,{u_{j+2}} + 496\,{u_{j+3}} - 307\,{u_{j+4}} + 104\,{u_{j+5}}}{8\,h^3} \nonumber \\
&& \mbox{} - \frac{15\,{u_{j+6}} }{8\,h^3} + \frac{29}{15}h^4 u^{(7)}(\xi)
\end{eqnarray}
\end{itemize}

\hspace{0.5cm}
\item Fourth derivative
\begin{itemize}
\item centered:
\begin{eqnarray}
u^{(4)}(x) &=& \frac{-{u_{j-3}} + 12\,{u_{j-2}} - 39\,{u_{j-1}} + 56\,{u_j} - 39\,{u_{j+1}} + 12\,{u_{j+2}} - {u_{j+3}} }{6\,h^4} \nonumber  \\
&& \mbox{} - \frac{7}{240}h^4u^{(8)}(\xi)
\end{eqnarray}

\item off-centered (1 point):
\begin{eqnarray}
u^{(4)}(x) &=& \frac{4\,{u_{j-2}} - 11\,{u_{j-1}} + 31\,{u_{j+1}} - 44\,{u_{j+2}} + 27\,{u_{j+3}} - 8\,{u_{j+4}} + {u_{j+5}}}{6\,h^4} \nonumber \\
&& \mbox{} + \frac{11}{80} h^4 u^{(8)}(\xi)
\end{eqnarray}

\item off-centered (2 point):
\begin{eqnarray}
u^{(4)}&=& \frac{21\,{u_{j-1}} - 112\,{u_j} + 255\,{u_{j+1}} - 324\,{u_{j+2}} +  251\,{u_{j+3}} - 120\,u_{j+4} }{6\,h^4} \nonumber \\
&& \mbox{}+\frac{33\,{u_{j+5}} - 4\,{u_{j+6}} }{6\,h^4} - \frac{127}{240}h^4 u^{(8)}(\xi)
\end{eqnarray}

\item off-centered (3 points):
\begin{eqnarray}
u^{(4)}&=& \frac{56\,{u_j} - 333\,{u_{j+1}} + 852\,{u_{j+2}} - 1219\,{u_{j+3}} + 1056\,{u_{j+4}} - 555\,{u_{j+5}} }{6\,h^4} \nonumber \\
&&\mbox{} + \frac{ 164\,u_{j+6}-  21\,u_{j+7}}{6\,h^4} - \frac{967}{240} h^4 u^{(8)}(\xi)
\end{eqnarray}
\end{itemize}

\end{itemize}

\bibliographystyle{apsrev}
\bibliography{articulo1}

\end{document}